\documentclass[lettersize,journal]{IEEEtran}
\usepackage{amsmath,amsfonts}
\usepackage{algorithmic}
\usepackage{algorithm}
\usepackage{array}
\usepackage[caption=false,font=normalsize,labelfont=sf,textfont=sf]{subfig}
\usepackage{textcomp}
\usepackage{stfloats}
\usepackage{url}
\usepackage{verbatim}
\usepackage{graphicx}
\usepackage{cite}

\usepackage[pscoord]{eso-pic}

\newcommand{\placetextbox}[3]{
  \setbox0=\hbox{#3}
  \AddToShipoutPictureFG*{
    \put(\LenToUnit{#1\paperwidth},\LenToUnit{#2\paperheight}){\vtop{{\null}\makebox[0pt][c]{#3}}}%
  }%
}%

\usepackage{bm}

\usepackage{multirow}

\usepackage[draft]{minted}
\usepackage{float}
\floatstyle{ruled}
\newfloat{code}{thp}{lop}
\floatname{code}{Code Block}

\usepackage{nomencl}

\newcommand{\imjl}{InfrastructureModels.jl}
\newcommand{\pmitdjl}{PowerModelsITD.jl}
\newcommand{\pmdjl}{PowerModelsDistribution.jl}
\newcommand{\pmjl}{PowerModels.jl}

\begin{document}

\title{Modeling and Rapid Prototyping of Integrated Transmission-Distribution OPF Formulations with PowerModelsITD.jl}

\author{Juan Ospina, David M. Fobes, Russell Bent, and Andreas W\"achter
\thanks{
(\textit{Corresponding author}: J. Ospina )

J. Ospina, D. Fobes, and R. Bent are with Los Alamos National Laboratory, Los Alamos, NM, USA (jjospina, dfobes, rbent@lanl.gov)

A. W\"achter is with the Northwestern University, Evanston, IL, USA (waechter@iems.northwestern.edu)

}
}

\maketitle

\begin{abstract}
Conventional electric power systems are composed of different unidirectional power flow stages of generation, transmission, and distribution, managed independently by transmission system and distribution system operators. However, as distribution systems increase in complexity due to the integration of distributed energy resources, coordination between transmission and distribution networks will be imperative for the optimal operation of the power grid. However, coupling models and formulations between transmission and distribution is non-trivial, in particular due to the common practice of modeling transmission systems as single-phase, and distribution systems as multi-conductor phase-unbalanced. To enable the rapid prototyping of power flow formulations, in particular in the modeling of the boundary conditions between these two seemingly incompatible data models, we introduce \pmitdjl, a free, open-source toolkit written in Julia for integrated transmission-distribution (ITD) optimization that leverages mature optimization libraries from the \imjl-ecosystem. The primary objective of the proposed framework is to provide baseline implementations of steady-state ITD optimization problems, while providing a common platform for the evaluation of emerging formulations and optimization problems. In this work, we introduce the nonlinear formulations currently supported in \pmitdjl, which include AC-polar, AC-rectangular, current-voltage, and a linear network transportation model. Results are validated using combinations of IEEE transmission and distribution networks.
\end{abstract}


\section*{Nomenclature}
\label{sect:nomen}
\addcontentsline{toc}{section}{Nomenclature}

\textit{Sets}
\begin{IEEEdescription}[\IEEEusemathlabelsep\IEEEsetlabelwidth{$----$}]
\item[$\mathcal{T}$] Belongs to transmission network.
\item[$\mathcal{D}$] Belongs to distribution network.
\item[$\mathcal{B}$] Set of boundary buses.
\item[$\Lambda$] Set of boundary links.
\item[$N$] Set of buses.
\item[$R$] Set of reference buses.
\item[$G$] Set of generators.
\item[$G_i$] Generator (Gen.) at bus $i$.
\item[$E, E_R$] Set of branches (forward and reverse).
\end{IEEEdescription}

\textit{Parameters}
\begin{IEEEdescription}[\IEEEusemathlabelsep\IEEEsetlabelwidth{$----$}]
\item[$\Re$] Real part.
\item[$\Im$] Imaginary part.
\item[$\Phi = a, b, c$] Multi-conductor phases.
\item[$\chi \rightarrow{\mathcal{T}},{\mathcal{D}}$]  Belongs to $\mathcal{T}$ or $\mathcal{D}$.
\item[$\beta^{^{\chi}}$] Boundary bus.
\item[$v_i^l$] Voltage lower bounds.
\item[$v_i^u$] Voltage upper bounds.
\item[$P_{g,k}^{^{\chi,l}}$] Gen.\ active power lower bounds.
\item[$P_{g,k}^{^{\chi,u}}$] Gen.\ active power upper bounds.
\item[$Q_{g,k}^{^{\chi,l}}$] Gen.\ reactive power lower bounds.
\item[$Q_{g,k}^{^{\chi,u}}$] Gen.\ reactive power upper bounds.
\item[$c_{2}, c_{1}, c_{0}$] Gen.\ cost components.
\item[$P_{d,i}^{\chi}$] Active power demand at bus $i$.
\item[$g_i^s$] Shunt conductance at bus $i$.
\item[$b_i^s$] Shunt susceptance at bus $i$.
\item[$p_{ij}^{\chi,l}$] Active power flow on line $(i,j)$ lower bounds.
\item[$p_{ij}^{\chi,u}$] Active power flow on line $(i,j)$ upper bounds.
\item[$q_{ij}^{\chi,l}$] Reactive power flow on line $(i,j)$ lower bounds.
\item[$q_{ij}^{\chi,u}$] Reactive power flow on line $(i,j)$ upper bounds.
\item[$\tau_{ij}$] Transformer tap ratio on line $(i,j)$.
\item[$\phi_{ij}$] Transformer angle on line $(i,j)$.
\item[$g_{ij}^E$] Conductance on line $(i,j)$.
\item[$b_{ij}^E$] Susceptance on line $(i,j)$.
\item[$b_{ij}^C$] Branch charging susceptance of line $(i,j)$.
\item[$x_{ij}$] Reactance on line $(i,j) \in E$.
\item[$\theta_{ij}^{\Delta l}$] Branch voltage angle difference lower bounds.
\item[$\theta_{ij}^{\Delta u}$] Branch voltage angle difference upper bounds.
\end{IEEEdescription}

\textit{Variables}

\textit{Transmission Variables}
\begin{IEEEdescription}[\IEEEusemathlabelsep\IEEEsetlabelwidth{$----$}]
\item[$P_{g,k}^{^\mathcal{T}}$] Gen.\ $k$ active power output.
\item[$Q_{g,k}^{^\mathcal{T}}$] Gen.\ $k$ reactive power output.
\item[$V_i^{^\mathcal{T}}$] Voltage magnitude at bus $i$.
\item[$\theta_i^{^\mathcal{T}}$] Voltage angle at bus $i$.
\item[$P_{ij}^{^\mathcal{T}}$] Active power flow on line $(i,j)$.
\item[$Q_{ij}^{^\mathcal{T}}$] Reactive power flow on line $(i,j)$.
\end{IEEEdescription}

\textit{Boundary Variables}
\begin{IEEEdescription}[\IEEEusemathlabelsep\IEEEsetlabelwidth{$----$}]
\item[$P_{\beta^{^\mathcal{T}}\beta^{^\mathcal{D}}}^{^\mathcal{T}}$] Active power flow from $\beta^{^\mathcal{T}}$ to $\beta^{^\mathcal{D}}$.
\item[$Q_{\beta^{^\mathcal{T}}\beta^{^\mathcal{D}}}^{^\mathcal{T}}$] Reactive power flow from $\beta^{^\mathcal{T}}$ to $\beta^{^\mathcal{D}}$.
\item[$P_{\beta^{^\mathcal{D}}\beta^{^\mathcal{T}}}^{{\mathcal{D},\varphi}}$] Active power flow from $\beta^{^\mathcal{D}}$ phase $\varphi$ to $\beta^{^\mathcal{T}}$.
\item[$Q_{\beta^{^\mathcal{D}}\beta^{^\mathcal{T}}}^{{\mathcal{D},\varphi}}$] Reactive power flow from $\beta^{^\mathcal{D}}$ phase $\varphi$ to $\beta^{^\mathcal{T}}$.
\end{IEEEdescription}

\textit{Distribution Variables}
\begin{IEEEdescription}[\IEEEusemathlabelsep\IEEEsetlabelwidth{$----$}]
\item[$P_{g,m}^{{\mathcal{D}},\varphi}$] Gen.\ $m$ active power output on phase $\varphi$.
\item[$Q_{g,m}^{{\mathcal{D}},\varphi}$] Gen.\ $m$ reactive power output on phase $\varphi$.
\item[$v_i^{\mathcal{D}, \varphi}$] Voltage magnitude at bus $i$ phase $\varphi$.
\item[$\theta_i^{\mathcal{D}, \varphi}$] Voltage angle at bus $i$ phase $\varphi$.
\item[$P_{ij}^{\mathcal{D}, \varphi}$] Active power flow on line $(i,j)$ phase $\varphi$.
\item[$Q_{ij}^{\mathcal{D}, \varphi}$] Reactive power flow on line $(i,j)$ phase $\varphi$.
\end{IEEEdescription}
\section{Introduction}
\label{sect:intro}

\subsection{Background}
\IEEEPARstart{T}{\lowercase{he}} unprecedented acceleration in the deployments of distributed energy resources (DERs), such as distributed generation (DG) and distributed storage (DS) systems, is increasing the complexity for the coordination and optimal control of these resources.

Traditionally, conventional power networks are composed of different stages of generation, transmission, and distribution managed independently by transmission system operators (TSOs) and distribution system operators (DSOs). For operation purposes, TSOs typically model distribution systems as bulk loads while DSOs model transmission systems as voltages sources connected at the substation. Due to the passive nature of traditional distribution systems, they are largely seen as a passive requester of energy for which the transmission system provides generation, voltage adjustments, and other controls. However, as power networks, more specifically distribution networks, become more active and complex due to the integration of DERs, intelligent control technologies, and the introduction of demand response (DR) programs, the assumption of the distribution system being just a load seen from the transmission system-side may no longer be valid and some problems could become computationally intractable when using existing methodologies. The introduction of advanced co-optimization methods and formulations, such as the ones proposed in this work, are designed to provide system planners and operators cross-domain visibility of system conditions and constraints that will enhance the coordination between transmission and distribution (T\&D) systems during normal operation and extreme scenarios. 

\subsection{\pmjl~and \pmdjl}
Fueled by the explosion in the number of different power flow nonconvex formulations, approximations, and relaxations for transmission systems being proposed by researchers, and the difficulty of their evaluation using a common platform, \textit{\pmjl}\footnote{https://github.com/lanl-ansi/PowerModels.jl} (PM) was introduced\cite{coffrin2018powermodels}. PM is a free and open-source software library designed to streamline the process of evaluating and comparing different power flow formulations based on a shared optimization problem specification (i.e., optimal power flow (OPF), power flow (PF), optimal transmission switching (OTS), etc.). 

\begin{figure}[h]
\centering
\includegraphics[width = 1.0\linewidth]{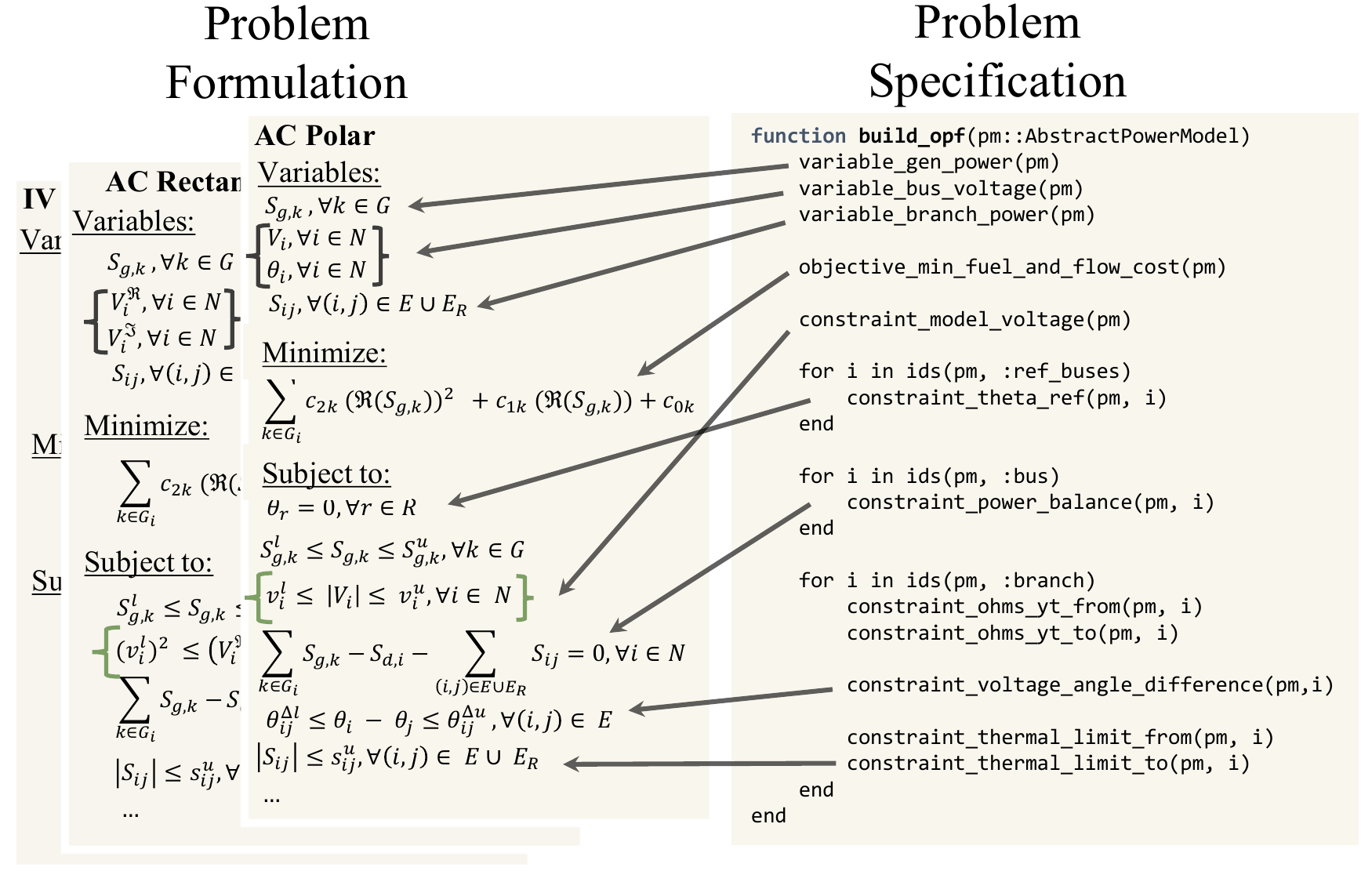}
\caption{\label{fig:core_design} OPF problem specification vs. formulations in \textit{PowerModels.jl}. Core design characterizing all packages belonging to the \textit{InfrastructureModels.jl} ecosystem.}
\end{figure}

Written in Julia and utilizing JuMP\footnote{https://jump.dev/}\cite{dunning2017jump}, PM was specifically designed for performing various quasi-steady-state optimizations of power transmission networks while utilizing a diverse set of power flow formulations applied to the same problem specification. This separation between optimization problem specification (e.g., OPF, PF, etc.) and problem formulation (e.g., AC polar, AC rectangular, etc.), as shown in Fig. \ref{fig:core_design}, is a key element that characterizes all packages that belong to the \textit{InfrastructureModels.jl} suite.

Some of the formulations, relevant to integrated transmission-distribution (ITD) optimization, available in PM include: AC polar, AC rectangular, current-voltage, and a linear transportation model, along with a broad selection of other relaxations and approximations.

Building on the success of \textit{\pmjl}, \textit{\pmdjl}\footnote{https://github.com/lanl-ansi/PowerModelsDistribution.jl} (PMD) \cite{fobes2020powermodelsdistribution} was introduced to address the same issues PM addresses while focusing on a baseline implementation of steady-state multi-conductor (e.g., three-phase) phase-unbalanced distribution network optimization problems. Similar to PM, PMD also provides many formulations such as the AC polar, AC rectangular, DC approximation, and SOC relaxations, among others. All these formulations can be evaluated using the same shared optimization problem specification (e.g., OPF, PF, etc.). A more detailed description of this package can be found in \cite{fobes2020powermodelsdistribution}.

\subsection{Introduction to \pmitdjl}
\label{sect:introtopmitd}
As previously described, PM and PMD excel at providing a robust and easy-to-use framework for evaluating transmission and distribution networks independently. Consequently, we have utilized them to form the basis of a new software framework capable of modeling optimization problems at the intersection of T\&D systems. In this work, we introduce a free, Julia-based, open-source, toolkit called \textit{\pmitdjl}\footnote{https://github.com/lanl-ansi/PowerModelsITD.jl}~(PMITD), designed for solving ITD power network optimization problems. {\pmitdjl} is part of the InfrastrutureModels.jl ecosystem written in the Julia programming language, whose merits for infrastructure modeling and optimization have been discussed previously in \cite{coffrin2018powermodels}.

This software framework already includes support for many formulations, where equivalents exist in both PM and PMD,
by creating the combined system cost function and the necessary boundary variables and constraints that interact at the T\&D boundary for those formulations. Because PMITD uses PM and PMD as the backend frameworks for generating the respective T\&D networks variables and constraints, PMITD needs only to contain capabilities to form the boundary interaction between the transmission and the distribution systems. Currently, the ITD formulations fully supported by PMITD are \textcolor{black}{[\textit{left is the formulation name used to model the transmission system in PM - right is the unbalanced formulation name used to model the distribution system(s) in PMD}]}:

\begin{enumerate}
    \item ACRPowerModel-ACRUPowerModel (AC rectangular)
    \item ACPPowerModel-ACPUPowerModel (AC polar)
    \item IVRPowerModel-IVRUPowerModel (IV rectangular)
    \item NFAPowerModel-NFAUPowerModel (Network-flow active power, i.e., transportation model)
\end{enumerate}

\subsection{Contributions}
\label{sect:contributions}

There exist some notable research gaps related to the study and evaluation of ITD power network optimization problems: 

\begin{itemize}
    \item To the authors’ knowledge, there is no readily available open-source software tool that allows conducting comprehensive ITD system studies based on optimization, while facilitating the assessment of the performance of different formulations. 
    
    \item The majority of the ongoing research focuses in decomposition models that separate the T\&D OPF models, contrary to our proposed approach. These are motivated assumptions such as the unwillingness from TSOs and DSOs to share and/or combine models \cite{coordinatedTanD} \cite{decentralizedacoptimal} and that centralized models for large-scale T\&D systems would not be scalable and are hard to solve due to its nonlinear and nonconvex nature \cite{decentralizedacoptimal}, \cite{tu2020two}.
    
    \item Many papers focus on a small number of formulations to compare and tend to focus on transmission-only or distribution-only systems; there is therefore a lack of studies that showcase a comparison of the different formulations available applied to the ITD OPF and PF problems.
    
\end{itemize}

To address these research gaps, we propose a framework that provides a baseline implementation of steady-state ITD network optimization problems. The intended user of the proposed framework is any researcher, student, or stakeholder that wants to experiment, research, study, evaluate, or solve steady-state ITD optimization problems using a diverse set of formulations and problem specifications. To this end, the paper describes the ITD OPF problem specification for a core collection of nonlinear formulations, and a simple linear approximation as a demonstration, while providing a common platform for their evaluation.  The contributions of this work can be summarized as follows:

\begin{itemize}

    \item A free open-source Julia-based software, called \textit{\pmitdjl}, designed for solving ITD power network optimization problems is proposed.
    
    \item A centralized ITD problem formulation algorithm is proposed as the baseline implementation. Boundary variables and constraints are defined to allow interfacing single-phase transmission system models with phase unbalanced distribution system model(s) based on a diverse collection of formulations. 
    
    \item A framework that allows users to perform rapid prototyping of experiments and evaluations based on a diverse set of problem specifications (OPF, PF, etc.) and formulations (AC rectangular, AC polar, etc.) is presented. Using this framework, users are able to evaluate the complexity and scalability of a specific test case by testing it against existing baseline formulations. Additionally, users have the ability to develop new problem specifications and/or ITD formulations by simply adding specific changes to the package based on the core design previously presented.
    
    \item Comparison studies between the different formulations available are conducted. In these studies, cost, runtime, and solver iterations are the main key metrics used to evaluate the functionality and performance of the proposed integrated approach. The integrated approach is evaluated against an \textit{independent} approach, where the T\&D systems are optimized independently.
\end{itemize}

The remainder of the paper is organized as follows. Section \ref{sect:relatedwork} presents a thorough literature review of the currently available methods, algorithms, and software applications designed to address the ITD problem. Section \ref{sect:itdopf} presents the proposed mathematical formulation of the centralized ITD-OPF problem specification for a diverse group of ITD formulations. In Section \ref{sect:usepmitd}, we introduce the \pmitdjl~ framework and give details regarding the currently supported formulations and problem specifications. Section \ref{sect:experimentsresults} presents the experimental setup and results of the case studies used to evaluate \pmitdjl. Finally, Section \ref{sect:conclusion} concludes the paper and enumerates directions for future work.

\section{Related Work}
\label{sect:relatedwork}

\subsection{ITD-OPF Problems}

Many researchers have focused on addressing the ITD-OPF problem. For instance, the authors in \cite{coordinatedTanD} present a coordinated T\&D AC-OPF formulation that is based on a main-problem/sub-problem structure and a heterogeneous decomposition algorithm (HGD). In the proposed model, the ITD problem is decomposed into a series of decoupled sub-problems designed to represent the transmission sub-problem and a group of distribution sub-problems. A similar approach is presented in \cite{decentralizedacoptimal}, where authors propose a decentralized approach based on a distribution-cost correction framework, that, in turn, is based on approximated distribution cost functions. The proposed approach formulates the transmission system-side problem using the AC-polar formulation and the distribution system-side problem using a second-order cone (SOC) branch flow (BF) model. 

The research presented in \cite{masterslave}, proposes a global power flow method that considers both T\&D system equations and solves the ITD problem using a main-problem/sub-problem/splitting iterative method with convergence guarantees that alleviate boundary variables mismatches. Complex power and voltage variables are exchanged between the transmission and the distribution system(s), while equivalent admittances, representing the radial distribution systems, are modeled as loads connected to the transmission system OPF model. In \cite{tu2020two}, researchers propose a scalable two-stage decomposition approach where the T\&D power network is decomposed into a primary network and multiple sub-networks. Each decomposed network has its own AC-OPF sub-problem that can be solved independently following a two-stage optimization approach. The main contribution is a smoothing technique that utilizes the properties of the barrier problem formulation, which naturally arise when the sub-problems are solved by a primal-dual interior-point algorithm. 

A hierarchical coordination approach based on Benders decomposition is presented in \cite{yuan2017hierarchical}. Here, the AC-OPF equations are convexified using SOC constraints. The review paper in \cite{givisiez2020review} presents a comprehensive review of the advantages, disadvantages, and challenges behind three core TSO-DSO coordination models (i.e., TSO-managed, DSO-managed, and TSO-DSO hybrid) and three techniques (i.e., Distributed, Hierarchical, and Centralized) identified by the authors.

Another class of studies where researchers propose decomposition algorithms for multi-area transmission systems have also been performed. For instance, in \cite{biskas2006decentralised}, a method for the decentralized solution of the OPF problem of large-scale interconnected transmission systems is proposed. In this approach, the overall problem is decomposed into independent OPF sub-problems, one for each area, and these sub-problems are coordinated through pricing mechanisms and iterative loops.
In \cite{arnold2007improvement}, methodology adjustments designed to divide the central OPF problem into sub-problems are proposed and solved using two coordination-decomposition approaches: a) \textit{adjustment at interface} and b) \textit{passing adjacent variables}.

T\&D modeling is also being addressed by co-simulation efforts aimed at creating platforms capable of accurately simulating T\&D systems. HELICS\cite{helics} is one example of a framework capable of simulating complex T\&D architectures; other co-simulation platforms and frameworks are summarized in \cite{cosim2}. Although these frameworks and platforms are capable of simulating large-scale power networks, enabling real-time analyses, they do not provide optimization-based frameworks for rapid prototyping of novel optimization formulations.

\subsection{OPF Formulations}

Besides the ITD-OPF studies summarized above, researchers have also tackled comparing the quality and performance of the different OPF formulations that exist for T\&D power networks. One example is \cite{BOLFEK2021106935}, where the capability of distribution networks to adjust their active and reactive power flows at the T\&D boundary by using three different formulations is explored using the AC-rect, an SOC-based relaxation, and an exact BF formulation. Similarly, the researchers in \cite{sereeter} compare four mathematical formulations of the OPF problem for transmission systems, i.e., the AC-polar, AC-rect, IV-rect, and IV-polar, by evaluating their performances based on computational burden, number of iterations, and number of nonzero elements in the Jacobian and Hessian matrices. These formulations are implemented in MATPOWER and use various solvers such as MATPOWER’s Interior Point Method (MIPS), KNITRO, and FMINCON. 

Other formulations, such as the backward-forward sweep (BFS), solved using Gurobi, have also been compared to the standard AC-polar method, solved using IPOPT, in terms of computational burden and quality of the solution when used to solve large-scale active power distribution networks \cite{mylonas2021comparison}. A similar study is conducted in \cite{lee2015performance}, where the authors conduct studies to compare various OPF algorithms applied to large-scale Korean power transmission system networks. 
A detailed feasibility comparison study, based on a mathematical analysis, of the DC and AC-OPF formulations is presented in \cite{dcnotfeasible}, concluding that DC-OPF solutions are rarely feasible, even when adjusting generation and accounting for losses. 

An OPF model designed for AC-DC transmission power networks is presented in \cite{hakanergun}. A novel relaxation is presented in \cite{qcrelaxation}, where researchers formulate a quadratic convex (QC) relaxation that imposes constraints that preserve stronger links between voltage variables, when compared to the SOC, semidefinite programming (SDP), and copper plate relaxations, via the use of convex envelopes. The authors perform detailed quality and runtime comparison studies. The surveys presented in \cite{fercstudies}, \cite{frank2012optimal1}, and \cite{frank2012optimal2} showcase extensive literature reviews regarding all currently available AC-OPF formulations, ranging from deterministic to non-deterministic approaches. However, these surveys focus heavily on formulations for transmission systems. 

While there are many open-source and commercial tools available to perform OPF and PF studies for either transmission or distribution systems, detailed reviews of which can be found in \cite{nguyen2021software} and \cite{sperstad2016optimal}, these tools focus on simulating either transmission-only or distribution-only power networks and do not provide capabilities to solve ITD power network optimization problems.

Based on the literature review conducted, notable research gaps have been identified as described in Section \ref{sect:contributions}. In the following sections, we present the proposed mathematical formulation of the ITD OPF problem specification and validate the performance and functionality of the proposed PMITD framework using a multitude of test cases.

\section{Integrated Transmission-Distribution (ITD) OPF Problem Specification}
\label{sect:itdopf}

The OPF problem can be characterized as a cost minimization problem with equality constraints in charge of enforcing Kirchhoff's current law, i.e., the power balance at each bus, and inequality constraints that represent the physical and stability limits on the power flows and generations. There are many formulations used to represent the power flow equations. Some of these formulations are the AC-polar and AC-rect formulations, the IV-rect formulation, relaxations such as SOC and SDP, and linear approximations such as LinDistFlow and the active-power-only network flow approximation transportation model (NFA). In this section, we provide a detailed description of the centralized ITD OPF problem specification based on the nonlinear AC-polar formulation. Subsequently, the respective boundary variables and constraints for different ITD formulations, i.e., AC-rect, IV-rect, and transportation model, are presented.

\subsection{AC-polar (ACP-ACPU) Formulation}

The AC-polar formulation is characterized for modeling the true physics of the AC power flow model by making use of the polar form of complex quantities while explicitly using sine and cosine functions in the power flow constraints of the problem. Variables such as $P_i/Q_i$, $V_i$, and $P_{ij}/Q_{ij}$ are introduced into the model to represent the active/reactive powers provided by the generators at bus $i$, the voltage at bus $i$, and the active/reactive power on line $(i,j) \in E \cup E_{R}$.

The OPF problem specification for the polar AC ITD model (ACP-ACPU) is largely based on the traditional AC-polar formulation \cite{Cain2012}.
Eq. (\ref{eq:acpopfitdcostfunc}) shows the objective function of the ITD problem, where the operating costs of all generators in both transmission and distribution systems are considered.  

\begin{equation}
\label{eq:acpopfitdcostfunc}
\begin{split}
   &\text{min~~} \bigg(\sum_{k \in G^{^\mathcal{T}}} c_{2k}(P_{g,k}^{\mathcal{T}})^2 + c_{1k}(P_{g,k}^{\mathcal{T}}) + c_{0k} \bigg) +\\ 
   &\bigg(\sum_{m \in G^{^\mathcal{D}}} c_{2m}(\sum_{\varphi \in \Phi} P_{g,m}^{\mathcal{D},\varphi})^2 + c_{1m}(\sum_{\varphi \in \Phi} P_{g,m}^{\mathcal{D},\varphi}) + c_{0m} \bigg)
\end{split}
\end{equation}

\noindent The ITD problem is subject to the following constraints:

\noindent \underline{\textit{Generation (Transmission)}}: Eq. (\ref{eq:transrefbus}) enforces the reference bus angle constraint and Eqs. (\ref{eq:transP})--(\ref{eq:transQ}) enforce the generators' power limits.

\vspace{-6mm}
\begin{gather}
    \label{eq:transrefbus}
    \theta_r^{^\mathcal{T}} = 0, \ \forall r \in R \\
    \label{eq:transP}
    P_{g,k}^{^{\mathcal{T},l}},  \leq P_{g,k}^{^\mathcal{T}} \leq P_{g,k}^{^{\mathcal{T},u}}, \ \ \ \forall k \in G^{^\mathcal{T}} \\
    \label{eq:transQ}
    Q_{g,k}^{^{\mathcal{T},l}},  \leq Q_{g,k}^{^\mathcal{T}} \leq Q_{g,k}^{^{\mathcal{T},u}}, \ \ \ \forall k \in G^{^\mathcal{T}}
\end{gather}

\noindent \underline{\textit{Power flow physics (Transmission)}}: The power flow physics, thermal limits, and bus power balances are described by the following set of constraints: 
\begin{gather}
    \label{eq:transbusVmag}
    v_i^l \leq |V_i| \leq v_i^u, \ \forall i \in N^{^\mathcal{T}}
\end{gather}
\begin{gather}
    \label{eq:transactivepowerflowij}
    \begin{split}
    P_{ij}^{^\mathcal{T}} &= \frac{1}{\tau_{ij}^2}g_{ij}^{E^{\mathcal{T}}}V_i^2 - \frac{1}{\tau_{ij}}V_i V_j(g_{ij}^{E^{\mathcal{T}}} cos(\theta_i - \theta_j - \phi_{ij}) \\
    & ...+b_{ij}^{E^{\mathcal{T}}}sin(\theta_i - \theta_j - \phi_{ij})), \ \ \ \forall (i,j) \in E^{^\mathcal{T}}\\\\
    \end{split}
    \\
     \label{eq:transactivepowerflowji}
    \begin{split}
    P_{ji}^{^\mathcal{T}} &= g_{ij}^{E^{\mathcal{T}}}V_j^2 - \frac{1}{\tau_{ij}}V_i V_j(g_{ij}^{E^{\mathcal{T}}} cos(\theta_j - \theta_i + \phi_{ij}) \\
    & ...+b_{ij}^{E^{\mathcal{T}}}sin(\theta_j - \theta_i + \phi_{ij})), \ \ \ \forall (i,j) \in E^{^\mathcal{T}}\\\\
    \end{split}\\
    \label{eq:transreactivepowerflowij}
    \begin{split}
    Q_{ij}^{^\mathcal{T}} &= -\frac{1}{\tau_{ij}^2}\Bigg(b_{ij}^{E^{\mathcal{T}}}+\frac{b_{ij}^{C}}{2}\Bigg)V_i^2 - \frac{1}{\tau_{ij}}V_i V_j(g_{ij}^{E^{\mathcal{T}}} cos(\theta_i \\ 
    & ... - \theta_j - \phi_{ij})-b_{ij}^{E^{\mathcal{T}}}sin(\theta_i - \theta_j - \phi_{ij})), \\
    \ \ \ ... &\forall (i,j) \in E^{^\mathcal{T}}\\\\
    \end{split}
\end{gather}
\begin{gather}
     \label{eq:transreactivepowerflowji}
    \begin{split}
    Q_{ji}^{^\mathcal{T}} &= -\Bigg(b_{ij}^{E^{\mathcal{T}}}+\frac{b_{ij}^{C}}{2}\Bigg)V_j^2 - \frac{1}{\tau_{ij}}V_i V_j(g_{ij}^{E^{\mathcal{T}}} cos(\theta_j \\ 
    &...- \theta_i + \phi_{ij})-b_{ij}^{E^{\mathcal{T}}}sin(\theta_j - \theta_i + \phi_{ij})), \ \ \ \\
    & ...\forall (i,j) \in E^{^\mathcal{T}}\\\\
    \end{split}\\
    \begin{split}
    \label{eq:transmissionacnoderealpowerbalance}
    &\sum_{k\in G_i^{^\mathcal{T}}} P_{g,k}^{^\mathcal{T}} - P_{d,i}^{^\mathcal{T}} - (V_i)^2g_{i}^{s} -  \!\!\!\!\!\!\!\!\!\!\!\!\sum_{(i,j)\in E^{^\mathcal{T}} \cup E_{R}^{^\mathcal{T}}}  \!\!\!\!\!\!\!\!\!\!\!\!P_{ij}^{^\mathcal{T}} \\
    &...-  \!\!\!\!\!\!\!\!\!\!\!\!\sum_{(i,\beta) \in \Lambda, \beta \in N^{^\mathcal{D}}\cap N^{^\mathcal{B}}} \!\!\!\!\!\!\!\!\!\!\!\! P_{i\beta}^{^\mathcal{T}} = 0, \ \ \ \forall i \in N^{^\mathcal{T}}\\\\
    \end{split}\\
    \begin{split}
    \label{eq:transmissionacnodereactivepowerbalance}
    &\sum_{k\in G_i^{^\mathcal{T}}} Q_{g,k}^{^\mathcal{T}} - Q_{d,i}^{^\mathcal{T}} - (V_i)^2b_{i}^{s} - \!\!\!\!\!\!\!\!\!\!\!\!\sum_{(i,j)\in E^{^\mathcal{T}} \cup E_{R}^{^\mathcal{T}}}  \!\!\!\!\!\!\!\!\!\!\!\!Q_{ij}^{^\mathcal{T}} \\  
    & ...-  \!\!\!\!\!\!\!\!\!\!\!\!\sum_{(i,\beta) \in \Lambda, \beta \in N^{^\mathcal{D}}\cap N^{^\mathcal{B}}}  \!\!\!\!\!\!\!\!\!\!\!\!Q_{i\beta}^{^\mathcal{T}} = 0, \ \ \ \forall i \in N^{^\mathcal{T}}\\
    \end{split}\\
     \label{eq:transactivepowerlinelimit}
    |P_{ij}| \leq p_{ij}^{^{^{\mathcal{T},u}}}, \ \ \ \forall(i,j)\in E^{^\mathcal{T}} \cup E_{R}^{^\mathcal{T}}\\
    \label{eq:transreactivepowerlinelimit}
    |Q_{ij}| \leq q_{ij}^{^{^{\mathcal{T},u}}}, \ \ \ \forall(i,j)\in E^{^\mathcal{T}} \cup E_{R}^{^\mathcal{T}}\\
    \label{eq:transbusangledifferences}
    \theta_{ij}^{\Delta l} \leq \theta_i - \theta_j \leq \theta_{ij}^{\Delta u}, \ \ \ \forall (i,j) \in E^{^\mathcal{T}}
\end{gather}

\noindent Eq. (\ref{eq:transbusVmag}) enforces the bus voltage magnitude limits. Eqs. (\ref{eq:transactivepowerflowij})--(\ref{eq:transactivepowerflowji}) compute the active power flow on lines $i,j \in E^{^\mathcal{T}}$, while Eqs. (\ref{eq:transreactivepowerflowij})--(\ref{eq:transreactivepowerflowji}) compute the line's reactive power flow. Eqs. (\ref{eq:transmissionacnoderealpowerbalance})--(\ref{eq:transmissionacnodereactivepowerbalance}) define the bus active and reactive power balance constraints. A new term that represents the active and reactive power flowing from the transmission to the distribution system(s) is added. This term is only used for buses that exist at the `boundary'. Eqs. (\ref{eq:transactivepowerlinelimit})--(\ref{eq:transbusangledifferences}) define the line's active and reactive power limits, and the bus angle differences, respectively.\\

\noindent \underline{\textit{T\&D power flow (Boundary)}}: The integration of the transmission and distribution system(s) at their respective boundaries is modeled by the following set of constraints:
\begin{gather}
    \label{eq:acpboundaryrealpowerequality}
    \sum_{\varphi \in \Phi} P_{\beta^{^\mathcal{D}}\beta^{^\mathcal{T}}}^{\mathcal{D},\varphi} +  P_{\beta^{^\mathcal{T}}\beta^{^\mathcal{D}}}^{^\mathcal{T}} = 0, \ \forall (\beta^{^\mathcal{T}},\beta^{^\mathcal{D}}) \in {\Lambda}\\
    \label{eq:acpboundaryreactivepowerequality}
    \sum_{\varphi \in \Phi} Q_{\beta^{^\mathcal{D}}\beta^{^\mathcal{T}}}^{\mathcal{D},\varphi} +  Q_{\beta^{^\mathcal{T}}\beta^{^\mathcal{D}}}^{^\mathcal{T}} = 0, \ \forall (\beta^{^\mathcal{T}},\beta^{^\mathcal{D}}) \in {\Lambda} \ \ \\
    \label{eq:acpboundaryvoltagemagnitudeAequality}
    V_{\beta^{^\mathcal{T}}} = v_{\beta^{^\mathcal{D}}}^{^{a}}, \ \forall (\beta^{^\mathcal{T}},\beta^{^\mathcal{D}}) \in {\Lambda} \ \\
    \label{eq:acpboundaryvoltagemagnitudeBequality}
    V_{\beta^{^\mathcal{T}}} = v_{\beta^{^\mathcal{D}}}^{^{b}}, \ \forall (\beta^{^\mathcal{T}},\beta^{^\mathcal{D}}) \in {\Lambda} \ \\
    \label{eq:acpboundaryvoltagemagnitudeCequality}
    V_{\beta^{^\mathcal{T}}} = v_{\beta^{^\mathcal{D}}}^{^{c}}, \ \forall (\beta^{^\mathcal{T}},\beta^{^\mathcal{D}}) \in {\Lambda} \ \\
    \label{eq:acpboundaryvoltageangleAequality}
    \theta_{\beta^{^\mathcal{T}}} = \theta_{\beta^{^\mathcal{D}}}^{^{a}}, \ \forall (\beta^{^\mathcal{T}},\beta^{^\mathcal{D}}) \in {\Lambda} \ \\
    \label{eq:acpboundaryvoltageangleBequality}
    \theta_{\beta^{^\mathcal{D}}}^{^{b}} = (\theta_{\beta^{^\mathcal{D}}}^{^{a}} -120^{\circ}),  \ \forall \beta^{^\mathcal{D}} \in N^{^\mathcal{B}} \cap  N^{^\mathcal{D}} \\
    \label{eq:acpboundaryvoltageangleCequality}
    \ \theta_{\beta^{^\mathcal{D}}}^{^{c}} = (\theta_{\beta^{^\mathcal{D}}}^{^{a}} +120^{\circ}), \ \forall \beta^{^\mathcal{D}} \in N^{^\mathcal{B}} \cap  N^{^\mathcal{D}}
\end{gather}    

\noindent Eqs. (\ref{eq:acpboundaryrealpowerequality})--(\ref{eq:acpboundaryreactivepowerequality}) equalize the active and reactive powers flowing at the boundary. The active and reactive power flowing from the transmission system boundary bus, $\beta^{^\mathcal{T}}$, to the distribution system boundary bus,  $\beta^{^\mathcal{D}}$, are equalized to the negative summation over all phases (i.e.,  phases $\Phi \in \{a,b,c\}$) of the powers flowing from the distribution system boundary bus to the transmission system boundary bus, where $\forall (\beta^{^\mathcal{T}},\beta^{^\mathcal{D}}) \in \Lambda$. Eqs. (\ref{eq:acpboundaryvoltagemagnitudeAequality})--(\ref{eq:acpboundaryvoltageangleCequality}) define the equality constraints related to the voltage magnitudes and angles of the boundary buses. These constraints allow the one-to-one mapping of the single-phase-to-three-phase boundary buses.

\noindent \underline{\textit{Generation (Distribution)}}: Eqs. (\ref{eq:distP})-(\ref{eq:distQ}) enforce the distributed generators' power limits.
\begin{gather}
      \label{eq:distP}
      \begin{bmatrix} P_{g,m}^{^{\mathcal{D},l,a}} \\  P_{g,m}^{^{\mathcal{D},l,b}} \\  P_{g,m}^{^{\mathcal{D},l,c}} \end{bmatrix}  \leq \begin{bmatrix} P_{g,m}^{^{\mathcal{D},a}} \\  P_{g,m}^{^{\mathcal{D},b}} \\  P_{g,m}^{^{\mathcal{D},c}} \end{bmatrix}  \leq  \begin{bmatrix} P_{g,m}^{^{\mathcal{D},u,a}} \\  P_{g,m}^{^{\mathcal{D},u,b}} \\  P_{g,m}^{^{\mathcal{D},u,c}} \end{bmatrix}, \ \ \ \forall m \in G^{^\mathcal{D}} \\
    \label{eq:distQ}
    \begin{bmatrix} Q_{g,m}^{^{\mathcal{D},l,a}} \\  Q_{g,m}^{^{\mathcal{D},l,b}} \\  Q_{g,m}^{^{\mathcal{D},l,c}} \end{bmatrix}  \leq \begin{bmatrix} Q_{g,m}^{^{\mathcal{D},a}} \\  Q_{g,m}^{^{\mathcal{D},b}} \\  Q_{g,m}^{^{\mathcal{D},c}} \end{bmatrix}  \leq  \begin{bmatrix} Q_{g,m}^{^{\mathcal{D},u,a}} \\  Q_{g,m}^{^{\mathcal{D},u,b}} \\  Q_{g,m}^{^{\mathcal{D},u,c}} \end{bmatrix}, \ \ \ \forall m \in G^{^\mathcal{D}}
\end{gather}

\noindent \underline{\textit{Power flow physics (Distribution)}}: The power flow physics, thermal limits, and node power balances are described by the following set of constraints: 
\vspace{-2mm}

\begin{gather}
    \label{eq:distbusVmag}
    \begin{bmatrix} v_i^{l,a} \\ v_i^{l,b} \\ v_i^{l,c} \end{bmatrix}  \leq \begin{bmatrix} |v_i^{a}| \\ |v_i^{b}| \\ |v_i^{c}| \end{bmatrix} \leq  \begin{bmatrix} v_i^{u,a} \\ v_i^{u,b} \\ v_i^{u,c} \end{bmatrix}, \ \forall i \in N^{^\mathcal{D}}\\
    \label{eq:distactivepowerflowij}
    \begin{split}
    P_{ij}^{{\mathcal{D},\varphi}} &= \frac{1}{(\tau_{ij}^{\varphi})^2}g_{ij}^{\varphi}(v_i^\varphi)^2 - \frac{1}{\tau_{ij}^{\varphi}}v_i^\varphi \sum_{\rho = a,b,c} v_j^{\rho}\bigg(g_{ij}^{\varphi \rho} cos(\theta_i^{\varphi}\\ 
    &...- \theta_j^{\rho} - \phi_{ij}^{\varphi \rho})+b_{ij}^{\varphi \rho}sin(\theta_i^{\varphi} - \theta_j^{\rho} - \phi_{ij}^{\varphi \rho})\bigg),\\
    &...\ \ \ \forall \varphi \in \Phi, \ \forall (i,j) \in E^{^\mathcal{D}}\\\\
    \end{split}
    \\
    \label{eq:distactivepowerflowji}
    \begin{split}
     P_{ji}^{\mathcal{D}, \varphi} &= g_{ij}^{\varphi \rho}(v_j^{\varphi})^2 - \frac{1}{\tau_{ij}^{\varphi}}v_i^{\varphi} \sum_{\rho=a,b,c} v_j^{\rho}\bigg(g_{ij}^{\varphi \rho} cos(\theta_j^{\varphi} \\
     &...- \theta_i^{\rho} + \phi_{ij}^{\varphi \rho})+b_{ij}^{\varphi \rho}sin(\theta_j^{\varphi} - \theta_i^{\rho} + \phi_{ij}^{\varphi \rho})\bigg), \\
     &...\ \ \ \forall \varphi \in \Phi, \ \forall (i,j) \in E^{^\mathcal{D}}\\\\
    \end{split}\\
    \label{eq:distreactivepowerflowij}
    \begin{split}
     Q_{ij}^{\mathcal{D},\varphi} &= -\frac{1}{(\tau_{ij}^{\varphi})^2}\Bigg(b_{ij}^{\varphi}+\frac{b_{ij}^{C,\varphi}}{2}\Bigg)(v_i^{\varphi})^2 \\
     &...- \frac{1}{\tau_{ij}^{\varphi}}v_i^{\varphi} \sum_{\rho=a,b,c} v_j^{\rho}\bigg(g_{ij}^{\varphi \rho} cos(\theta_i^{\varphi} - \theta_j^{\rho} - \phi_{ij}^{\varphi \rho})\\
     &...-b_{ij}^{\varphi \rho}sin(\theta_i^{\varphi} - \theta_j^{\rho} - \phi_{ij}^{\varphi \rho})\bigg),\\
     &...\ \forall \varphi \in \Phi, \ \forall (i,j) \in E^{^\mathcal{D}}\\\\
     \end{split}\\
     \label{eq:distreactivepowerflowji}
     \begin{split}
     Q_{ji}^{\mathcal{D},\varphi} &= -\Bigg(b_{ij}^{\varphi}+\frac{b_{ij}^{C,\varphi}}{2}\Bigg)(v_j^{\varphi})^2 \\
     &...- \frac{1}{\tau_{ij}^{\varphi}}v_i^{\varphi} \sum_{\rho=a,b,c} v_j^{\rho}\bigg(g_{ij}^{\varphi \rho} cos(\theta_j^{\varphi} - \theta_i^{\rho} + \phi_{ij}^{\varphi \rho})\\
     &...-b_{ij}^{\varphi \rho}sin(\theta_j^{\varphi} - \theta_i^{\rho} + \phi_{ij}^{\varphi \rho})\bigg), \\
     &...\ \forall \varphi \in \Phi, \ \forall (i,j) \in E^{^\mathcal{D}}\\\\
     \end{split}
\end{gather}
\begin{gather}
     \begin{split}
    \label{eq:distributionacnoderealpowerbalance}
    &\sum_{m\in G_i^{^\mathcal{D}}} \sum_{\varphi \in \Phi} P_{g,m}^{\mathcal{D},\varphi} - \sum_{\varphi \in \Phi} P_{d,i}^{\mathcal{D},\varphi} - \sum_{\varphi \in \Phi} (v_i^{\varphi})^2 g_{i}^{s,\varphi} \\
    &...- \sum_{(i,j)\in E^{^\mathcal{D}} \cup E_{R}^{^\mathcal{D}}} \sum_{\varphi \in \Phi} P_{ij}^{\mathcal{D},\varphi}  \\
    &...- \sum_{(i,\beta) \in \Lambda, \beta \in N^{^\mathcal{T}}\cap N^{^\mathcal{B}}} \sum_{\varphi \in \Phi} P_{i\beta}^{\mathcal{D},\varphi} = 0, \ \ \ \forall i \in N^{^\mathcal{D}}\\\\
    \end{split}\\
    \begin{split}
    \label{eq:distributionacnodereactivepowerbalance}
    &\sum_{m\in G_i^{^\mathcal{D}}} \sum_{\varphi \in \Phi} Q_{g,m}^{\mathcal{D},\varphi} - \sum_{\varphi \in \Phi} Q_{d,i}^{\mathcal{D},\varphi} - \sum_{\varphi \in \Phi} (v_i^{\varphi})^2 b_{i}^{s,\varphi} \\
    &...- \sum_{(i,j)\in E^{^\mathcal{D}} \cup E_{R}^{^\mathcal{D}}} \sum_{\varphi \in \Phi} Q_{ij}^{\mathcal{D},\varphi} \\ 
    &...- \sum_{(i,\beta) \in \Lambda, \beta \in N^{^\mathcal{T}}\cap N^{^\mathcal{B}}} \sum_{\varphi \in \Phi} Q_{i\beta}^{\mathcal{D},\varphi} = 0, \ \ \ \forall i \in N^{^\mathcal{D}}\\\\
    \end{split}\\
    \label{eq:distactivepowerlinelimit}
    \begin{bmatrix} |P_{ij}^{^{{\mathcal{D},a}}}| \\  |P_{ij}^{^{{\mathcal{D},b}}}| \\ |P_{ij}^{^{{\mathcal{D},c}}}|  \end{bmatrix} \leq \begin{bmatrix} p_{ij}^{^{{\mathcal{D},u,a}}} \\   p_{ij}^{^{{\mathcal{D},u,b}}} \\   p_{ij}^{^{{\mathcal{D},u,c}}} \end{bmatrix}, \ \forall(i,j)\in E^{^\mathcal{D}} \cup E_{R}^{^\mathcal{D}}\\
    \label{eq:distreactivepowerlinelimit}
    \begin{bmatrix} |Q_{ij}^{^{{\mathcal{D},a}}}| \\  |Q_{ij}^{^{{\mathcal{D},b}}}| \\ |Q_{ij}^{^{{\mathcal{D},c}}}|  \end{bmatrix} \leq \begin{bmatrix} q_{ij}^{^{{\mathcal{D},u,a}}} \\   q_{ij}^{^{{\mathcal{D},u,b}}} \\   q_{ij}^{^{{\mathcal{D},u,c}}} \end{bmatrix}, \ \forall(i,j)\in E^{^\mathcal{D}} \cup E_{R}^{^\mathcal{D}}\\
    \begin{split}\\
    \label{eq:distbusangledifferences}
    \begin{bmatrix} \theta_{ij}^{\Delta l,a} \\  \theta_{ij}^{\Delta l,b} \\ \theta_{ij}^{\Delta l,c}  \end{bmatrix} \!\! \leq \!\!  \begin{bmatrix} \theta_{i}^{^{\mathcal{D},a}} \\  \theta_{i}^{^{\mathcal{D},b}} \\ \theta_{i}^{^{\mathcal{D},c}}  \end{bmatrix} \!\! - \!\! \begin{bmatrix} \theta_{j}^{^{\mathcal{D},a}} \\  \theta_{j}^{^{\mathcal{D},b}} \\ \theta_{j}^{^{\mathcal{D},c}}  \end{bmatrix} \!\! \leq \!\! \begin{bmatrix} \theta_{ij}^{\Delta u,a} \\  \theta_{ij}^{\Delta u,b} \\ \theta_{ij}^{\Delta u,c}  \end{bmatrix}, \forall (i,j) \in E^{^\mathcal{D}}\\
     \end{split}
\end{gather}

\noindent Eq. (\ref{eq:distbusVmag}) enforces the voltage magnitude limits at each phase in bus $i$. Eqs. (\ref{eq:distactivepowerflowij})--(\ref{eq:distreactivepowerflowji}) compute the active and reactive power flow on lines $i,j \in E^{^\mathcal{D}}$ and phase $\varphi \in \Phi$. Eqs. (\ref{eq:distributionacnoderealpowerbalance})--(\ref{eq:distributionacnodereactivepowerbalance}) define the nodes' active and reactive power balance constraints. A new term that represents the powers flowing from the distribution to the transmission system is added to these constraints (similar to the process done in transmission). Eqs. (\ref{eq:distactivepowerlinelimit})--(\ref{eq:distbusangledifferences}) define the line's active and reactive power limits, and the nodes angle differences, respectively.

\begin{figure}[h]
\centering
\includegraphics[width = 0.44\textwidth]{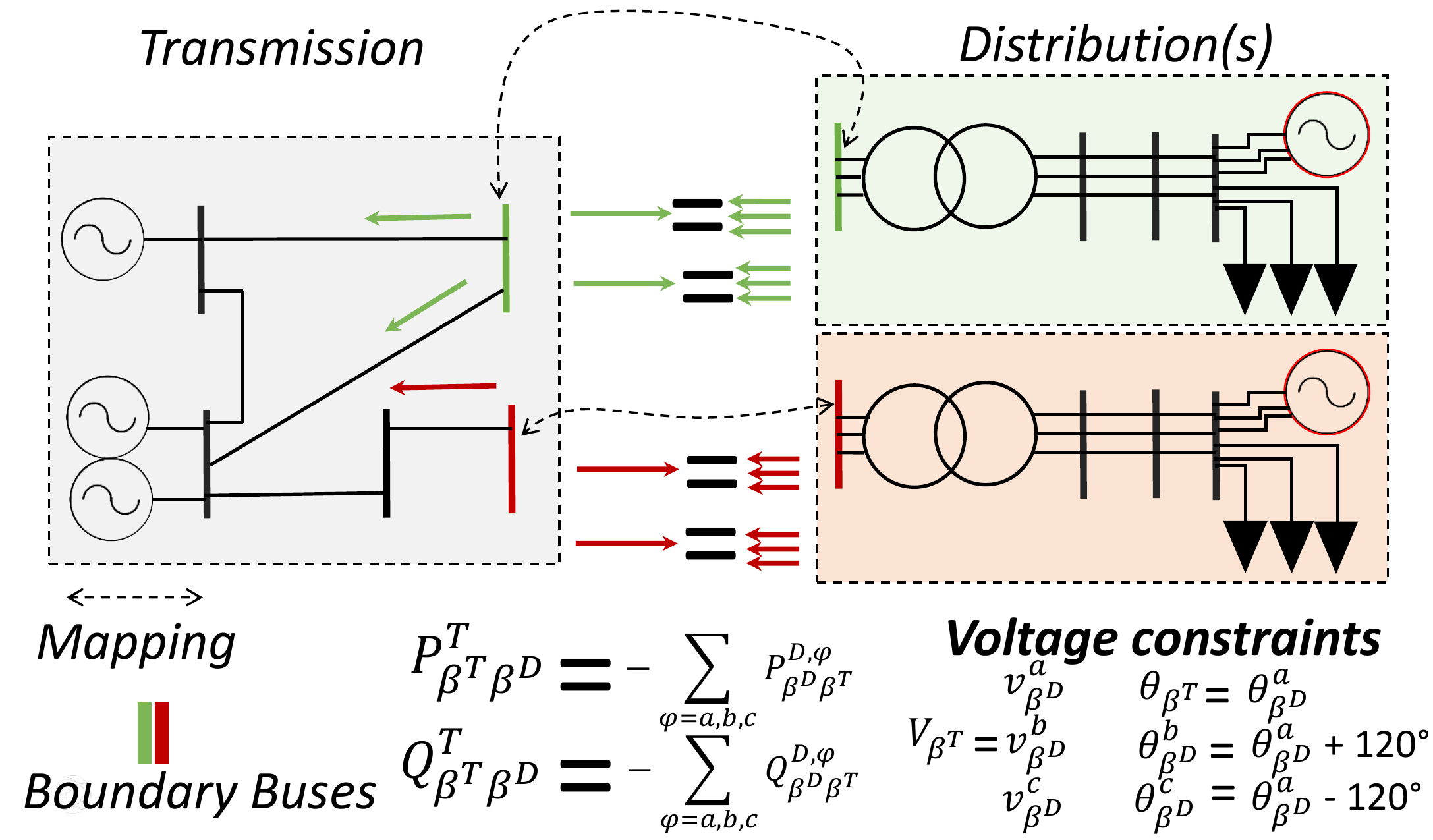}
\caption{\label{fig:TandDmap} Mapping of single-phase transmission system with three-phase distribution system(s) in the ACP-ACPU formulation.}
\end{figure}

As seen in the presented problem specification, the constraints that make up the transmission and distribution system sections are, for the most part, the same constraints defined in the traditional AC-OPF formulation. The main adjustments made are reflected in Eqs. (\ref{eq:transmissionacnoderealpowerbalance})--(\ref{eq:transmissionacnodereactivepowerbalance}),  (\ref{eq:distributionacnoderealpowerbalance})--(\ref{eq:distributionacnodereactivepowerbalance}), and (\ref{eq:acpboundaryrealpowerequality})--(\ref{eq:acpboundaryvoltageangleCequality}), which allow the integration between the transmission and the distribution system(s). Fig. \ref{fig:TandDmap} depicts the ACP-ACPU formulation mapping between the single-phase transmission system buses and the three-phase distribution system nodes at the T\&D boundaries. As seen in the figure, the voltage constraints enforce the mathematical mapping of the boundary buses, and the power flow constraints ensure that the power coming from the transmission system boundary bus is equal to the power flowing towards the rest of the distribution system. 

Due to space limitation and to avoid repeating equations and/or constraints, other formulations will be presented based on the boundary constraints only.

\subsection{AC-rect (ACR-ACRU) Boundary Formulation}

This subsection presents the boundary variables and constraints for the AC-rect formulation where transmission and distribution both use rectangular voltage coordinates. 
The boundary constraints are presented below. The superscript $\Re$ refers to the real part of the variable and $\Im$ refers to the imaginary part of the variable. Eqs. (\ref{eq:acrboundaryrealpowerequality})--(\ref{eq:acrboundaryreactivepowerequality}) equalize the active and reactive powers flowing at the boundary, while Eqs. (\ref{eq:acrboundaryvoltagemagnitudeAequality})--(\ref{eq:acrboundaryvoltagemagnitudeCequality}) and (\ref{eq:acrboundaryvoltageangleAequality})--(\ref{eq:acrboundaryvoltageangleCequality}) define  the  equality  constraints  related  to  the voltage magnitudes and angles of the boundary buses in rectangular coordinates, respectively.

\begin{gather}
    \label{eq:acrboundaryrealpowerequality}
    \sum_{\varphi \in \Phi} P_{\beta^{^\mathcal{D}}\beta^{^\mathcal{T}}}^{\mathcal{D},\varphi} +  P_{\beta^{^\mathcal{T}}\beta^{^\mathcal{D}}}^{^\mathcal{T}} = 0, \ \forall (\beta^{^\mathcal{T}},\beta^{^\mathcal{D}}) \in \Lambda\\
    \label{eq:acrboundaryreactivepowerequality}
    \sum_{\varphi \in \Phi} Q_{\beta^{^\mathcal{D}}\beta^{^\mathcal{T}}}^{\mathcal{D},\varphi} +  Q_{\beta^{^\mathcal{T}}\beta^{^\mathcal{D}}}^{^\mathcal{T}} = 0, \ \forall (\beta^{^\mathcal{T}},\beta^{^\mathcal{D}}) \in \Lambda
\end{gather}

\begin{gather}
    \label{eq:acrboundaryvoltagemagnitudeAequality}
    \Big({V^\Re_{\beta^{^\mathcal{T}}}}\Big)^2 \!\!\!\! + \! \Big({V^\Im_{\beta^{^\mathcal{T}}}}\Big)^2 \!\!\!\! = \!\Big(v_{\beta^{^\mathcal{D}}}^{^{a,\Re}}\Big)^2 \!\!\!\!+\! \Big(v_{\beta^{^\mathcal{D}}}^{^{a,\Im}}\Big)^2,\forall (\beta^{^\mathcal{T}},\beta^{^\mathcal{D}}) \in \Lambda \ \\
    \label{eq:acrboundaryvoltagemagnitudeBequality}
    \Big({V^\Re_{\beta^{^\mathcal{T}}}}\Big)^2 \!\!\!\!+\! \Big({V^\Im_{\beta^{^\mathcal{T}}}}\Big)^2 \!\!\!\!=\! \Big(v_{\beta^{^\mathcal{D}}}^{^{b,\Re}}\Big)^2 \!\!\!\!+\! \Big(v_{\beta^{^\mathcal{D}}}^{^{b,\Im}}\Big)^2,\forall (\beta^{^\mathcal{T}},\beta^{^\mathcal{D}}) \in \Lambda \ \\
    \label{eq:acrboundaryvoltagemagnitudeCequality}
   \Big({V^\Re_{\beta^{^\mathcal{T}}}}\Big)^2 \!\!\!\!+\! \Big({V^\Im_{\beta^{^\mathcal{T}}}}\Big)^2 \!\!\!\!=\! \Big(v_{\beta^{^\mathcal{D}}}^{^{c,\Re}}\Big)^2 \!\!\!\!+\! \Big(v_{\beta^{^\mathcal{D}}}^{^{c,\Im}}\Big)^2,\forall (\beta^{^\mathcal{T}},\beta^{^\mathcal{D}}) \in \Lambda\\ 
   \begin{split}
   \\
    \label{eq:acrboundaryvoltageangleAequality}
    \Big({V^\Im_{\beta^{^\mathcal{T}}}}\Big) = \Big(v_{\beta^{^\mathcal{D}}}^{^{a,\Im}}\Big),\ \ \ \forall (\beta^{^\mathcal{T}},\beta^{^\mathcal{D}}) \in \Lambda \ \\
    \end{split}
    \\
    \begin{split}
    \label{eq:acrboundaryvoltageangleBequality}
    \Big(v_{\beta^{^\mathcal{D}}}^{^{b,\Im}}\Big) = tan\Bigg(atan\bigg(\frac{v_{\beta^{^\mathcal{D}}}^{^{a,\Im}}}{v_{\beta^{^\mathcal{D}}}^{^{a,\Re}}} \bigg) &-120^{\circ} \Bigg) v_{\beta^{^\mathcal{D}}}^{^{b,\Re}}, \\
    &... \ \forall \beta^{^\mathcal{D}} \in N^{^\mathcal{B}} \cap  N^{^\mathcal{D}}
    \end{split}
    \\
    \begin{split}
    \label{eq:acrboundaryvoltageangleCequality}
    \Big(v_{\beta^{^\mathcal{D}}}^{^{c,\Im}}\Big) = tan\Bigg(atan\bigg(\frac{v_{\beta^{^\mathcal{D}}}^{^{a,\Im}}}{v_{\beta^{^\mathcal{D}}}^{^{a,\Re}}} \bigg) &+120^{\circ} \Bigg) v_{\beta^{^\mathcal{D}}}^{^{c,\Re}}, \\
    &... \ \forall \beta^{^\mathcal{D}} \in N^{^\mathcal{B}} \cap  N^{^\mathcal{D}}
    \end{split}
\end{gather}

\subsection{IV-rect (IVR-IVRU) Boundary Formulation}

This subsection presents the boundary variables and constraints for the rectangular current-voltage formulation. The full IV-rect formulation for unbalanced power systems can be found in \cite{geth2020current}. The boundary constraints for the described formulation are presented below. The symbol $\Re(\cdot)$ refers to the real part of the variable and $\Im(\cdot)$ refers to the imaginary part of the variable. Eqs. (\ref{eq:ivrboundaryrealcurrentequality})--(\ref{eq:ivrboundaryimaginarycurrentequality}) equalize the active and reactive powers flowing at the boundary (computed using the real and imaginary voltage and current), while Eqs. (\ref{eq:ivrboundaryvoltagemagnitudeAequality})--(\ref{eq:ivrboundaryvoltagemagnitudeCequality}) and (\ref{eq:ivrboundaryvoltageangleAequality})--(\ref{eq:ivrboundaryvoltageangleCequality}) define  the  equality  constraints  related  to  the voltage magnitudes and angles of the boundary buses in rectangular coordinates.

\begin{gather}
    \begin{split}
    \label{eq:ivrboundaryrealcurrentequality}
    &{V^\Re_{\beta^{^\mathcal{T}}}} \Re\Big(I_{\beta^{^\mathcal{T}}\beta^{^\mathcal{D}}}^{^\mathcal{T}}\Big) + {V^\Im_{\beta^{^\mathcal{T}}}} \Im\Big(I_{\beta^{^\mathcal{T}}\beta^{^\mathcal{D}}}^{^\mathcal{T}}\Big) = \\
    &... \!\!-\!\!\Bigg[\sum_{\varphi \in \Phi} \Bigg( \Big(v_{\beta^{^\mathcal{D}}}^{^{\varphi,\Re}}\Big) \Re\Big(I_{\beta^{^\mathcal{D}}\beta^{^\mathcal{T}}}^{\mathcal{D},\varphi}\Big) \!\!+\!\! \Big(v_{\beta^{^\mathcal{D}}}^{^{\varphi,\Im}}\Big) \Im\Big(I_{\beta^{^\mathcal{D}}\beta^{^\mathcal{T}}}^{\mathcal{D},\varphi}\Big) \Bigg) \Bigg],\\
    &... \ \forall (\beta^{^\mathcal{T}},\beta^{^\mathcal{D}}) \in \Lambda\\\\
    \end{split}\\
    \begin{split}
    \label{eq:ivrboundaryimaginarycurrentequality}
    &{V^\Im_{\beta^{^\mathcal{T}}}} \Re\Big(I_{\beta^{^\mathcal{T}}\beta^{^\mathcal{D}}}^{^\mathcal{T}}\Big) - {V^\Re_{\beta^{^\mathcal{T}}}} \Im\Big(I_{\beta^{^\mathcal{T}}\beta^{^\mathcal{D}}}^{^\mathcal{T}}\Big) = \\
    &... \!\!-\!\! \Bigg[\sum_{\varphi \in \Phi} \Bigg( \Big(v_{\beta^{^\mathcal{D}}}^{^{\varphi,\Im}}\Big) \Re\Big(I_{\beta^{^\mathcal{D}}\beta^{^\mathcal{T}}}^{\mathcal{D},\varphi}\Big) \!\!-\!\! \Big(v_{\beta^{^\mathcal{D}}}^{^{\varphi,\Re}}\Big) \Im\Big(I_{\beta^{^\mathcal{D}}\beta^{^\mathcal{T}}}^{\mathcal{D},\varphi}\Big) \Bigg) \Bigg], \\
    &... \ \forall (\beta^{^\mathcal{T}},\beta^{^\mathcal{D}}) \in \Lambda\\\\
    \end{split}\\
    \label{eq:ivrboundaryvoltagemagnitudeAequality}
    \Big({V^\Re_{\beta^{^\mathcal{T}}}}\Big)^2 \!\!\!\!+\! \Big({V^\Im_{\beta^{^\mathcal{T}}}}\Big)^2 \!\!\!\!=\! \Big(v_{\beta^{^\mathcal{D}}}^{^{a,\Re}}\Big)^2 \!\!\!\!+\! \Big(v_{\beta^{^\mathcal{D}}}^{^{a,\Im}}\Big)^2,\forall (\beta^{^\mathcal{T}},\beta^{^\mathcal{D}}) \in \Lambda \ \\
    \label{eq:ivrboundaryvoltagemagnitudeBequality}
    \Big({V^\Re_{\beta^{^\mathcal{T}}}}\Big)^2 \!\!\!\!+\! \Big({V^\Im_{\beta^{^\mathcal{T}}}}\Big)^2 \!\!\!\!=\! \Big(v_{\beta^{^\mathcal{D}}}^{^{b,\Re}}\Big)^2 \!\!\!\!+\! \Big(v_{\beta^{^\mathcal{D}}}^{^{b,\Im}}\Big)^2,\forall (\beta^{^\mathcal{T}},\beta^{^\mathcal{D}}) \in \Lambda \\  
    \label{eq:ivrboundaryvoltagemagnitudeCequality}
   \Big({V^\Re_{\beta^{^\mathcal{T}}}}\Big)^2 \!\!\!\!+\! \Big({V^\Im_{\beta^{^\mathcal{T}}}}\Big)^2 \!\!\!\!=\! \Big(v_{\beta^{^\mathcal{D}}}^{^{c,\Re}}\Big)^2 \!\!\!\!+\! \Big(v_{\beta^{^\mathcal{D}}}^{^{c,\Im}}\Big)^2,\forall (\beta^{^\mathcal{T}},\beta^{^\mathcal{D}}) \in \Lambda
\end{gather}
\begin{gather}
   \begin{split}
    \label{eq:ivrboundaryvoltageangleAequality}
    \Big({V^\Im_{\beta^{^\mathcal{T}}}}\Big) = \Big(v_{\beta^{^\mathcal{D}}}^{^{a,\Im}}\Big), \ \ \ \ \forall (\beta^{^\mathcal{T}},\beta^{^\mathcal{D}}) \in \Lambda \ \\
    \end{split}
    \\
    \begin{split}
    \label{eq:ivrboundaryvoltageangleBequality}
    \Big(v_{\beta^{^\mathcal{D}}}^{^{b,\Im}}\Big) = tan\Bigg(atan\bigg(\frac{v_{\beta^{^\mathcal{D}}}^{^{a,\Im}}}{v_{\beta^{^\mathcal{D}}}^{^{a,\Re}}} \bigg) &-120^{\circ} \Bigg) v_{\beta^{^\mathcal{D}}}^{^{b,\Re}}, \\
    &... \ \forall \beta^{^\mathcal{D}} \in N^{^\mathcal{B}} \cap  N^{^\mathcal{D}}\ \\
    \end{split}\\
    \begin{split}
    \label{eq:ivrboundaryvoltageangleCequality}
    \Big(v_{\beta^{^\mathcal{D}}}^{^{c,\Im}}\Big) = tan\Bigg(atan\bigg(\frac{v_{\beta^{^\mathcal{D}}}^{^{a,\Im}}}{v_{\beta^{^\mathcal{D}}}^{^{a,\Re}}} \bigg) &+120^{\circ} \Bigg) v_{\beta^{^\mathcal{D}}}^{^{c,\Re}}, \\
    &... \ \forall \beta^{^\mathcal{D}} \in N^{^\mathcal{B}} \cap  N^{^\mathcal{D}}
    \end{split}
\end{gather}

\subsection{Transportation model (NFA-NFAU) Boundary Formulation}

In the transportation model formulation, the only required boundary constraint, Eq. (\ref{eq:nfaboundaryrealpowerequality}), is the active power constraint at the boundary, since no voltage or reactive power variables exist. 
\begin{gather}
    \label{eq:nfaboundaryrealpowerequality}
    \sum_{\varphi \in \Phi} P_{\beta^{^\mathcal{D}}\beta^{^\mathcal{T}}}^{\mathcal{D},\varphi} +  P_{\beta^{^\mathcal{T}}\beta^{^\mathcal{D}}}^{^\mathcal{T}} = 0, \ \forall (\beta^{^\mathcal{T}},\beta^{^\mathcal{D}}) \in \Lambda
\end{gather}

\section{Using PowerModelsITD}
\label{sect:usepmitd}

In this section, the core design approach used in PMITD is presented.  The problem specification, formulation, and algorithmic flexibilites of the software package are explored. These flexibilities emerge from the use of the core design approach employed in PM and PMD (see Fig. \ref{fig:core_design}), where each power optimization problem specification (e.g., PF, OPF) is designed as a problem with well-defined semantics for a large set of formulations (e.g., AC rectangular, AC polar, etc.) adopting a specific coordination algorithm.

\subsection{Abstract Problem Specification}
Currently in PMITD, there are two main problem definitions: 1) Optimal Power Flow for ITD (OPFITD) and 2) Power Flow for ITD (PFITD). However, the package is designed to accommodate custom user-defined problem definitions through functional programming and multiple dispatch; core features of the Julia language. Code Block \ref{code:opfitdproblempart1} show a typical example of the OPFITD problem definition. Due to space limitation, the declaration of some constraints and variables have been replaced with the symbol (...). The sections related to the ITD formulation have been left verbatim. As seen, first the transmission and distribution models are separated and saved into independent model variables (i.e., {\texttt{pm\_model}} and {\texttt{pmd\_model}}). Every variable and constraint for each model will be applied to its respective model. Then, the variables for each model are instantiated (e.g., voltages, branch flows, generators, storage, etc.). It is important to note that to instantiate these variables, the respective PM or PMD library is used. The {\texttt{PM.}} and/or {\texttt{PMD.}} references refer to functions that belong to the respective software library. After the variables for the independent T\&D models are created, the specific PMITD boundary variables are instantiated. Then, constraints for the independent and boundary systems are applied, where the power balance of the T\&D buses need to be updated according to the OPFITD problem formulation. Finally, an objective function, in our case a standard minimum fuel cost objective, is added to the problem specification.

\begin{code}[h]
\caption{Problem specification for OPFITD}
\label{code:opfitdproblempart1}
\begin{minted}{julia}
function build_opfitd(pmitd::AbstractPowerModelITD) 
    pm_model = ... # Transmission model
    pmd_model = ... # Distribution model

    # PM(Transmission) Variables
    PM.variable_bus_voltage(pm_model)
    ...

    # PMD(Distribution) Variables
    PMD.variable_mc_bus_voltage(pmd_model)
    ...

    # PMITD (Boundary) Variables
    variable_boundary_power(pmitd)

    # --- PM(Transmission) Constraints ---
    PM.constraint_model_voltage(pm_model)
    ...

    # --- PMD(Distribution) Constraints ---
    PMD.constraint_mc_model_voltage(pmd_model)
    ...

    # --- PMITD-related Constraints -------
    for i in ids(pmitd, :boundary)
        constraint_boundary_power(pmitd, i)
        constraint_boundary_voltage_
                            magnitude(pmitd, i)
        constraint_boundary_voltage_
                            angle(pmitd, i)       
    end

    # ---- Transmission Power Balance ---
    boundary_buses = Vector{Int64}() 
    for i in PM.ids(pm_model, :bus)
        for j in ids(pmitd, :boundary)
            constraint_transmission_power_
                        balance_boundary(pmitd, i, 
                                j, boundary_buses)
        end
        if !(i in boundary_buses)
            PM.constraint_power_
                            balance(pm_model, i)
        end
    end

    # ---- Distribution Power Balance ---
    for i in PMD.ids(pmd_model, :bus)
        for j in ids(pmitd, :boundary)
            constraint_distribution_power_
                    balance_boundary(pmitd, i, 
                            j, boundary_buses)
        end
        if !(i in boundary_buses)
            PMD.constraint_mc_power_
                            balance(pmd_model, i)
        end
    end
    
    # --- PMITD Cost Functions --------    
    objective_itd_min_fuel_cost(pmitd)

end
\end{minted}
\end{code}

\subsection{Abstract Formulations}

As described in Section \ref{sect:intro}, the design approach taken allows the problem specification (e.g., OPFITD) to be separated from the abstract model formulation (e.g., AC-polar, which is denoted as ACP-ACPU, where ACP is the transmission formulation from PM, and ACPU is the phase-unbalanced distribution formulation from PMD). In essence, abstract formulations are defined for each model currently supported in PMITD, characterized by their representation of the electrical physics in a certain variable space (e.g., power-voltage, current-voltage, etc.), with a respective choice of coordinates (e.g., rectangular or polar). It should be noted that PMITD includes other experimental formulation pairings, such as AC and First-order Taylor (ACR-FOTRU, ACP-FOTPU) and AC and Forward-backward sweep (ACR-FBSUBF). However, these formulations have not been included in the results due to their experimental nature.

The combination of the generic abstract problem specification with the mathematical formulation of the model produces a fully specified mathematical program, encoded as a JuMP model, that can be solved with an appropriate JuMP-compatible solver. In terms of formulation flexibility, new formulations, for either the transmission or distribution system, can be added by simply defining the corresponding representative variables and constraints. The user only needs to define the new ITD formulation type and the boundary constraints for the new formulation in PMITD.

\subsection{Algorithmic Flexibility}

In terms of ITD coordination algorithms, the only ITD algorithm currently implemented in PMITD is the one presented in Section \ref{sect:itdopf}. However, this does not mean that this is the only algorithm PMITD is capable of supporting. Users that would like to develop new ITD coordination algorithms based on other approaches, such as decomposition, can write their own problem specification functions that define their desired algorithmic process.

\begin{algorithm}[H]
 \caption{\textit{Independent} ITD Coordination Algorithm}
 \begin{algorithmic}[1]
 \label{alg:algo1}
 \renewcommand{\algorithmicrequire}{\textbf{Input: }}
 \renewcommand{\algorithmicensure}{\textbf{Output:}}
 \REQUIRE \(Trans.\ Data\), \(Dist.\ Data\), $\Lambda$
 \\ \textit{Initialization}: \(ds\_runtimes\)=[...],\(ds\_costs\)=[...],\(ds\_P\)=[...], \(ds\_Q\)=[...]
  \FOR {\textbf{each} \textit{dist. system} ($ds$) in \(Dist.\ Data\)}
  \STATE Parse $ds$ data
  \STATE Limit upper bounds of DGs ($pg_{ub}$) in $ds$ for reserves
  \STATE Solve $ds$ OPF $\rightarrow$ \(ds\_sol\)
  \STATE \(ds\_P\)[...] $\leftarrow$ \(ds\_sol\) slack bus gen. P sum in all $\Phi$
  \STATE \(ds\_Q\)[...] $\leftarrow$ \(ds\_sol\) slack bus gen. Q sum in all $\Phi$
  \STATE \(ds\_costs\)[...] $\leftarrow$ \(ds\_sol\) cost
  \STATE \(ds\_runtimes\)[...] $\leftarrow$ \(ds\_sol\) runtime
  \ENDFOR
  \STATE  \(total\_dist\_cost\) = sum(\(ds\_costs\))
  \STATE \(total\_dist\_runtime\) = sum(\(ds\_runtimes\))
  \STATE Parse \textit{trans. system} ($ts$) data in \(Trans.\ Data\)
  \FOR {\textbf{each} ($\beta^{\mathcal{T}},\beta^{\mathcal{D}}$) in $\Lambda$}
    \STATE $P$ load at \(\beta^{\mathcal{T}} \in ts\) = \(ds\_P\)[$\beta^{\mathcal{D}} \in ds$]
    \STATE $Q$ load at \(\beta^{\mathcal{T}} \in ts\) = \(ds\_Q\)[$\beta^{\mathcal{D}} \in ds$]
  \ENDFOR
  \STATE Solve $ts$ OPF $\rightarrow$ \(ts\_sol\)
  \STATE Obtain objective cost from \(ts\_sol\)$\rightarrow$ \(ts\_cost\)
  \STATE Obtain solver time from \(ts\_sol\) $\rightarrow$ \(ts\_runtime\)
  \STATE \(total\_cost\) = \(ts\_cost\) + \(total\_dist\_cost\)
  \STATE \(total\_runtime\) = \(ts\_runtime\) + \(total\_dist\_runtime\)
 \end{algorithmic}
 \end{algorithm}

To demonstrate the simplicity of the process, Algorithm \ref{alg:algo1} presents the pseudocode for an `\textit{independent}' coordination ITD algorithm that can be implemented using PMITD. It is important to remark that this is a very simple ITD coordination algorithm and does not represent the state-of-the-art. This algorithm is mainly used for illustrative purposes. Also, note that the objective of this paper is not to test or evaluate all ITD coordination algorithms in existence nor to compare them under the same platform.

\subsection{Data Formats}

Preserving the file format compatibility in PM and PMD, PMITD allows the creation of a variety of large-scale realistic power networks via many standard file formats, such as MATPOWER (.m) and PSS/E (.raw), for transmission systems, and OpenDSS (.dss), for distribution system models. It should be noted that many other proprietary file formats are supported via conversion tools, such as DiTTo \cite{dittoreference}, that enable data exchange between a large variety of industry-dataset formats and open-source formats. \textcolor{black}{The data format for defining the boundaries (i.e., boundary linking file) between T\&D systems is the JSON format. (.json). This file contains information about the boundary links that connect the transmission system with the specific distribution system(s). In this file, the user defines the \textit{ids} of the buses that need to be mapped between the transmission and distribution systems.}

\subsection{How to use \pmitdjl}
One of the key features maintained in PMITD, similar to PM and PMD, is an intuitive user interface. As seen in Code Block \ref{code:pmitdusage}, to run an AC-polar OPFITD model the user only needs to: 1) load the transmission system, distribution system(s), and boundary linking files, 2) define the type of PMITD problem (in this case, we are using an AC-polar formulation for both the transmission and distribution systems), and 3) finally, use the function {\texttt{solve\_opfitd}}, which will build and solve the OPFITD model. \textcolor{black}{The building process is composed by three primary steps. First, all the data is parsed into a dictionary that contains all the mappings. Then, the mathematical problem specification is instantiated using the data parsed. Finally, the instantiated model is optimized using the provided optimizer.}

\begin{code}[]
\caption{Running AC-polar OPFITD with 1 distribution system.}
\label{code:pmitdusage}
\begin{minted}{julia}

using PowerModelsITD
import Ipopt
ipopt = Ipopt.Optimizer

pm_file = "transmission/case5_withload.m"
pmd_file = "distribution/case3_balanced.dss"
boundary_file = "boundary/case5_case3_bal.json"
pmitd_type = NLPowerModelITD{ACPPowerModel,
                            ACPUPowerModel}
                            
result = solve_opfitd(pm_file, pmd_file, 
                    boundary_file, pmitd_type, ipopt)
\end{minted}
\end{code}

It is important to note that the current version of PMITD supports passing multiple distribution system files. For cases where multiple distribution systems need to be connected to the transmission system (at different load buses), additional boundary links can be easily added to the PMITD boundary links JSON file (i.e., {\texttt{boundary\_file}}). PMITD checks for the existence and status of the provided buses and will display an error message to the user if any of them does not exist or is disabled. The multiple distribution system files (i.e., \texttt{pmd\_file}) need to be passed into the function {\texttt{solve\_opfitd}} as a vector of files.

\section{Results}
\label{sect:experimentsresults}

In this section, various experiments are conducted to evaluate the performance and demonstrate the value of using PMITD.

\subsection{Independent vs. ITD Optimizations}
\label{sect:experiment1}

\begin{figure}[t]
    \centering
    \subfloat[]{
     \includegraphics[width=0.9\linewidth]{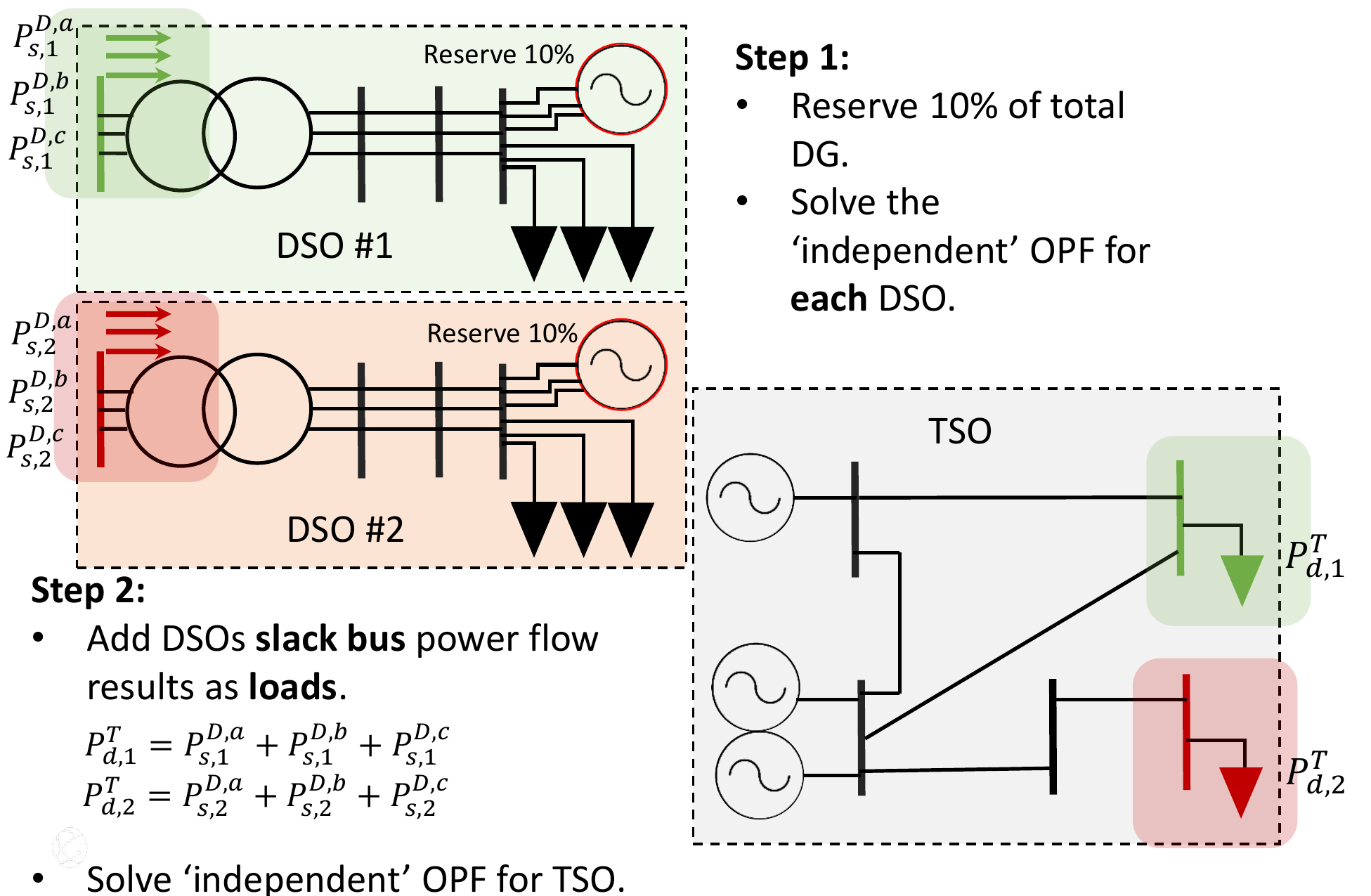}
     \label{fig:indonly}
    }\\
    \subfloat[]{
     \includegraphics[width=0.9\linewidth]{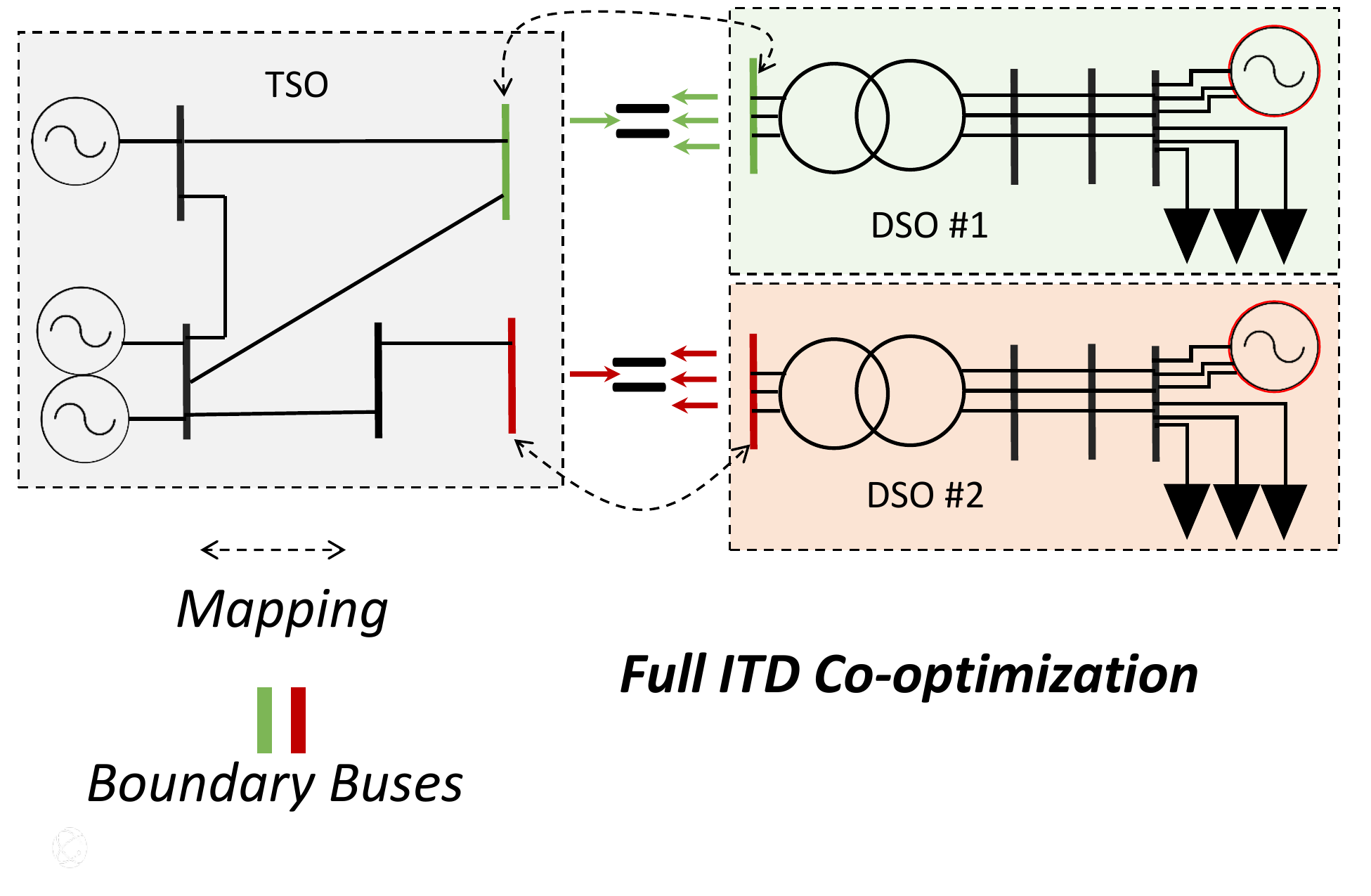}
     \label{fig:itdonly}
    }
    \caption[CR]{(a) \textit{Independent} vs. (b) \textit{ITD} optimization scenarios. Reactive power ($Q$) is not included in the figure, but it is included in the optimization processes.}
    \label{fig:indvsitd}
\end{figure}

In this experiment, the main objective is to evaluate the performance of PMITD against an \textit{independent} version of the test case where the transmission system and the distribution feeders(s) are run independently (i.e., without any integration) using PM and PMD, respectively. More specifically, in the independent scenarios, the distribution systems are optimized independently, assuming that DSOs want to maximize their distributed generation (DG) usage, but cautiously reserving 10\% of the total DG available in the system. On the other hand, in the integrated (i.e., PMITD) scenarios, the full use of the DG is allowed since coordination can be achieved due to the full integration of the systems. Fig. \ref{fig:indvsitd} shows a graphical description of the scenarios considered.

\subsubsection{Test Cases}
It should be noted that all distribution feeder models used here are Kron reduced. Based on the scenarios discussed, four test cases are defined:

\begin{enumerate}
    \item \textbf{Case5-Case3 with 1 DG}: A modified version of the PJM 5-bus system is used as the transmission system. The modification is realized by adding a new load bus connected to bus \#5. The distribution system is connected at the new load bus (bus \#5). The distribution system consists of a modified version of the IEEE 4 Node Test Feeder where the loads and line capacities are increased, and one 600 kW DG is added at bus \#4.
    \item \textbf{Case118-Case3 $\times$5 dist.\ systems with 1 DG}: In this test case, the IEEE 118-bus test system is used as the transmission system. Five distribution systems are connected at buses \#2, \#7, \#14, \#28, and \#44. The distribution systems consist of modified versions of the IEEE 4 Node Test Feeder where the loads and line capacities are increased, and one DG, with different power ratings, is added at bus \#4.
    \item \textbf{Case500-Case30 $\times$5 dist.\ systems with 1 DG}: In this test case, the IEEE PGLib synthetic 500-bus test system is used as the transmission system. Five distribution systems are connected at buses \#33, \#45, \#53, \#150, and \#178. The distribution systems consist of modified versions of the IEEE 30-bus test system where the system is made multi-conductor (three-phase) and one DG, rated at 40 MW, is added at bus \#B2.
    \item \textbf{Case500-CaseLV $\times$5 dist.\ systems with 3 DG}: In this test case, the IEEE PGLib synthetic 500-bus test system is used as the transmission system. Five distribution systems are connected at buses \#16, \#323, \#337, \#114, and \#366. The distribution systems consist of modified versions of the IEEE LVTestCase (an European Low-Voltage test feeder) system where the loads and line capacities are increased, and three DGs (rated at 100 kW) are added at buses \#835, \#539, and \#619.
\end{enumerate}

All the described test cases are included in \pmitdjl\footnote{https://github.com/lanl-ansi/PowerModelsITD.jl/blob/main/test/data}. Table \ref{tab:nodesedges} shows the total number of nodes and edges for each one of the test cases presented. Note that the `Totals' are calculated based on the number of nodes/edges in the transmission network plus the number of nodes/edges in each distribution network. In cases with 5 distribution systems (i.e., \textbf{Case 2, 3, and 4}), the total number of distribution systems' nodes/edges is equal to the number of nodes/edges per distribution system $\times5$, considering that the same system topology is used for all 5 distribution systems. Additionally, note that that the term \textit{node} is equivalent to the term \textit{bus} in the single-phase transmission system model, but they are not necessarily equivalent in the multiconductor distribution system model, since one \textit{bus} may have 3, 2, or 1 \textit{node(s)}, depending on the number of phases. IPOPT v3.13.4~\cite{wachter2006implementation} using the default linear solver MUMPS, is the solver used for all the problems presented here. The PMITD version used to run the test cases presented is v0.7.5.

\begin{table}[h]
\setlength{\tabcolsep}{1.2pt}
\caption{Total number of nodes and edges in networks used for test cases.}
\label{tab:nodesedges}
\centering
\begin{tabular}{||c|c|c|c|c|c|c||}
\hline \hline
\multicolumn{1}{|l|}{} & \multicolumn{2}{c|}{\textbf{Transmission}} & \multicolumn{2}{c|}{\textbf{Distribution}} & \multicolumn{2}{c|}{\textbf{Total}} \\ \hline 
\textbf{Test Case} & \textbf{|N|} & \textbf{|E|} & \textbf{|N|} & \textbf{|E|} & \textbf{|N|} & \textbf{|E|} \\ \hline \hline
\textbf{Case 1} & 6 & 7 & 12 & 3 & 18 & 10 \\ \hline
\textbf{Case 2} & 118 & 186 & 12 & 3 & 178 & 201 \\ \hline
\textbf{Case 3} & 500 & 733 & 90 & 41 & 950 & 938 \\ \hline
\textbf{Case 4} & 500 & 733 & 2724 & 907 & 14120 & 5268 \\ \hline \hline
\end{tabular}
\end{table}

\begin{table*}[h]
\setlength{\tabcolsep}{0.8pt}
\caption{Results for test case 1: case5-case3 with 1 DG}
\label{tab:results_case1}
\centering
\begin{tabular}{|c|cccccc|ccc|ccc|}
\hline
 & \multicolumn{6}{c|}{\textbf{Independent}} & \multicolumn{3}{c|}{\textbf{ITD}} & \multicolumn{3}{c|}{\multirow{2}{*}{\textbf{Differences}}} \\ \cline{1-10}
 & \multicolumn{3}{c|}{\textbf{PM}} & \multicolumn{3}{c|}{\textbf{PMD}} & \multicolumn{3}{c|}{\textbf{PMITD}} & \multicolumn{3}{c|}{} \\ \hline
\textbf{Formulation} & \multicolumn{1}{c|}{\textbf{\$/hr}} & \multicolumn{1}{c|}{\textbf{Time(s)}} & \multicolumn{1}{c|}{\textbf{Iterations}} & \multicolumn{1}{c|}{\textbf{\$/hr}} & \multicolumn{1}{c|}{\textbf{Time(s)}} & \textbf{Iterations} & \multicolumn{1}{c|}{\textbf{\$/hr}} & \multicolumn{1}{c|}{\textbf{Time(s)}} & \textbf{Iterations} & \multicolumn{1}{c|}{\textbf{\$/hr}} & \multicolumn{1}{c|}{\textbf{Time(s)}} & \textbf{Iterations} \\ \hline
\textbf{ACP-ACPU} & \multicolumn{1}{c|}{17756.0373} & \multicolumn{1}{c|}{0.0233} & \multicolumn{1}{c|}{22} & \multicolumn{1}{c|}{14.0401} & \multicolumn{1}{c|}{0.0178} & 7 & \multicolumn{1}{c|}{17770.0356} & \multicolumn{1}{c|}{0.0727} & 26 & \multicolumn{1}{c|}{0.0418} & \multicolumn{1}{c|}{-0.0316} & 3 \\ \hline
\textbf{ACR-ACRU} & \multicolumn{1}{c|}{17756.0373} & \multicolumn{1}{c|}{0.0285} & \multicolumn{1}{c|}{22} & \multicolumn{1}{c|}{14.0401} & \multicolumn{1}{c|}{0.0402} & 20 & \multicolumn{1}{c|}{17770.0356} & \multicolumn{1}{c|}{0.0652} & 26 & \multicolumn{1}{c|}{0.0418} & \multicolumn{1}{c|}{0.0036} & 16 \\ \hline
\textbf{IVR-IVRU} & \multicolumn{1}{c|}{17756.0374} & \multicolumn{1}{c|}{0.0252} & \multicolumn{1}{c|}{20} & \multicolumn{1}{c|}{14.0401} & \multicolumn{1}{c|}{0.0502} & 22 & \multicolumn{1}{c|}{17770.0357} & \multicolumn{1}{c|}{0.0757} & 29 & \multicolumn{1}{c|}{0.0418} & \multicolumn{1}{c|}{-0.0003} & 13 \\ \hline
\textbf{NFA-NFAU} & \multicolumn{1}{c|}{14534.2997} & \multicolumn{1}{c|}{0.0066} & \multicolumn{1}{c|}{14} & \multicolumn{1}{c|}{14.0401} & \multicolumn{1}{c|}{0.0064} & 7 & \multicolumn{1}{c|}{14548.0998} & \multicolumn{1}{c|}{0.0105} & 16 & \multicolumn{1}{c|}{0.2400} & \multicolumn{1}{c|}{0.0025} & 5 \\ \hline
\end{tabular}
\end{table*}

\begin{table*}[h]
\setlength{\tabcolsep}{0.8pt}
\caption{Results for test case 2: case118-case3 $\times$5 distribution systems with 1 DG.}
\label{tab:results_case2}
\centering
\begin{tabular}{|c|cccccc|ccc|ccc|}
\hline
 & \multicolumn{6}{c|}{\textbf{Independent}} & \multicolumn{3}{c|}{\textbf{ITD}} & \multicolumn{3}{c|}{\multirow{2}{*}{\textbf{Differences}}} \\ \cline{1-10}
 & \multicolumn{3}{c|}{\textbf{PM}} & \multicolumn{3}{c|}{\textbf{PMD}} & \multicolumn{3}{c|}{\textbf{PMITD}} & \multicolumn{3}{c|}{} \\ \hline
\textbf{Formulation} & \multicolumn{1}{c|}{\textbf{\$/hr}} & \multicolumn{1}{c|}{\textbf{Time(s)}} & \multicolumn{1}{c|}{\textbf{Iterations}} & \multicolumn{1}{c|}{\textbf{\$/hr}} & \multicolumn{1}{c|}{\textbf{Time(s)}} & \textbf{Iterations} & \multicolumn{1}{c|}{\textbf{\$/hr}} & \multicolumn{1}{c|}{\textbf{Time(s)}} & \textbf{Iterations} & \multicolumn{1}{c|}{\textbf{\$/hr}} & \multicolumn{1}{c|}{\textbf{Time(s)}} & \textbf{Iterations} \\ \hline
\textbf{ACP-ACPU} & \multicolumn{1}{c|}{94571.9597} & \multicolumn{1}{c|}{0.3401} & \multicolumn{1}{c|}{27} & \multicolumn{1}{c|}{64.7825} & \multicolumn{1}{c|}{0.0817} & 35 & \multicolumn{1}{c|}{94636.5198} & \multicolumn{1}{c|}{0.5523} & 28 & \multicolumn{1}{c|}{0.2224} & \multicolumn{1}{c|}{-0.1304} & 34 \\ \hline
\textbf{ACR-ACRU} & \multicolumn{1}{c|}{94571.9597} & \multicolumn{1}{c|}{0.3138} & \multicolumn{1}{c|}{28} & \multicolumn{1}{c|}{64.7825} & \multicolumn{1}{c|}{0.1780} & 98 & \multicolumn{1}{c|}{94636.5198} & \multicolumn{1}{c|}{0.5224} & 30 & \multicolumn{1}{c|}{0.2224} & \multicolumn{1}{c|}{-0.0305} & 96 \\ \hline
\textbf{IVR-IVRU} & \multicolumn{1}{c|}{94571.9597} & \multicolumn{1}{c|}{0.5835} & \multicolumn{1}{c|}{32} & \multicolumn{1}{c|}{64.7825} & \multicolumn{1}{c|}{0.2306} & 110 & \multicolumn{1}{c|}{94636.5199} & \multicolumn{1}{c|}{0.9002} & 33 & \multicolumn{1}{c|}{0.2223} & \multicolumn{1}{c|}{-0.0861} & 109 \\ \hline
\textbf{NFA-NFAU} & \multicolumn{1}{c|}{90893.1829} & \multicolumn{1}{c|}{0.0180} & \multicolumn{1}{c|}{15} & \multicolumn{1}{c|}{64.7825} & \multicolumn{1}{c|}{0.0250} & 35 & \multicolumn{1}{c|}{90947.8941} & \multicolumn{1}{c|}{0.0341} & 17 & \multicolumn{1}{c|}{10.0712} & \multicolumn{1}{c|}{0.0090} & 33 \\ \hline
\end{tabular}
\end{table*}

\begin{table*}[h]
\setlength{\tabcolsep}{0.8pt}
\caption{Results for test case 3: case500-case30 $\times$5 distribution systems with 1 DG.}
\label{tab:results_case3}
\centering
\begin{tabular}{|c|cccccc|ccc|ccc|}
\hline
 & \multicolumn{6}{c|}{\textbf{Independent}} & \multicolumn{3}{c|}{\textbf{ITD}} & \multicolumn{3}{c|}{\multirow{2}{*}{\textbf{Differences}}} \\ \cline{1-10}
 & \multicolumn{3}{c|}{\textbf{PM}} & \multicolumn{3}{c|}{\textbf{PMD}} & \multicolumn{3}{c|}{\textbf{PMITD}} & \multicolumn{3}{c|}{} \\ \hline
\textbf{Formulation} & \multicolumn{1}{c|}{\textbf{\$/hr}} & \multicolumn{1}{c|}{\textbf{Time(s)}} & \multicolumn{1}{c|}{\textbf{Iterations}} & \multicolumn{1}{c|}{\textbf{\$/hr}} & \multicolumn{1}{c|}{\textbf{Time(s)}} & \textbf{Iterations} & \multicolumn{1}{c|}{\textbf{\$/hr}} & \multicolumn{1}{c|}{\textbf{Time(s)}} & \textbf{Iterations} & \multicolumn{1}{c|}{\textbf{\$/hr}} & \multicolumn{1}{c|}{\textbf{Time(s)}} & \textbf{Iterations} \\ \hline
\textbf{ACP-ACPU} & \multicolumn{1}{c|}{470979.6190} & \multicolumn{1}{c|}{1.9290} & \multicolumn{1}{c|}{42} & \multicolumn{1}{c|}{180.0201} & \multicolumn{1}{c|}{0.8338} & 65 & \multicolumn{1}{c|}{469573.4476} & \multicolumn{1}{c|}{5.1104} & 45 & \multicolumn{1}{c|}{1586.1914} & \multicolumn{1}{c|}{-2.3476} & 62 \\ \hline
\textbf{ACR-ACRU} & \multicolumn{1}{c|}{470979.6191} & \multicolumn{1}{c|}{1.8811} & \multicolumn{1}{c|}{43} & \multicolumn{1}{c|}{180.0201} & \multicolumn{1}{c|}{3.0656} & 170 & \multicolumn{1}{c|}{469573.4478} & \multicolumn{1}{c|}{6.4275} & 55 & \multicolumn{1}{c|}{1586.1914} & \multicolumn{1}{c|}{-1.4808} & 158 \\ \hline
\textbf{IVR-IVRU} & \multicolumn{1}{c|}{470979.6192} & \multicolumn{1}{c|}{2.6622} & \multicolumn{1}{c|}{47} & \multicolumn{1}{c|}{180.0201} & \multicolumn{1}{c|}{2.0838} & 170 & \multicolumn{1}{c|}{469573.4479} & \multicolumn{1}{c|}{8.4899} & 52 & \multicolumn{1}{c|}{1586.1914} & \multicolumn{1}{c|}{-3.7439} & 165 \\ \hline
\textbf{NFA-NFAU} & \multicolumn{1}{c|}{450006.9826} & \multicolumn{1}{c|}{0.1180} & \multicolumn{1}{c|}{32} & \multicolumn{1}{c|}{180.0201} & \multicolumn{1}{c|}{0.0537} & 30 & \multicolumn{1}{c|}{449297.2305} & \multicolumn{1}{c|}{0.3989} & 34 & \multicolumn{1}{c|}{889.7721} & \multicolumn{1}{c|}{-0.2271} & 28 \\ \hline
\end{tabular}
\end{table*}

\begin{table*}[h]
\setlength{\tabcolsep}{0.8pt}
\caption{Results for test case 4: case500-caseLV $\times$5 distribution systems with 3 DG.}
\label{tab:results_case4}
\centering
\begin{tabular}{|c|cccccc|ccc|ccc|}
\hline
 & \multicolumn{6}{c|}{\textbf{Independent}} & \multicolumn{3}{c|}{\textbf{ITD}} & \multicolumn{3}{c|}{\multirow{2}{*}{\textbf{Differences}}} \\ \cline{1-10}
 & \multicolumn{3}{c|}{\textbf{PM}} & \multicolumn{3}{c|}{\textbf{PMD}} & \multicolumn{3}{c|}{\textbf{PMITD}} & \multicolumn{3}{c|}{} \\ \hline
\textbf{Formulation} & \multicolumn{1}{c|}{\textbf{\$/hr}} & \multicolumn{1}{c|}{\textbf{Time(s)}} & \multicolumn{1}{c|}{\textbf{Iterations}} & \multicolumn{1}{c|}{\textbf{\$/hr}} & \multicolumn{1}{c|}{\textbf{Time(s)}} & \textbf{Iterations} & \multicolumn{1}{c|}{\textbf{\$/hr}} & \multicolumn{1}{c|}{\textbf{Time(s)}} & \textbf{Iterations} & \multicolumn{1}{c|}{\textbf{\$/hr}} & \multicolumn{1}{c|}{\textbf{Time(s)}} & \textbf{Iterations} \\ \hline
\textbf{ACP-ACPU} & \multicolumn{1}{c|}{451045.6962} & \multicolumn{1}{c|}{1.8797} & \multicolumn{1}{c|}{39} & \multicolumn{1}{c|}{49.9517} & \multicolumn{1}{c|}{15.2863} & 55 & \multicolumn{1}{c|}{451093.0292} & \multicolumn{1}{c|}{62.6413} & 50 & \multicolumn{1}{c|}{2.6186} & \multicolumn{1}{c|}{-45.4752} & 44 \\ \hline
\textbf{ACR-ACRU} & \multicolumn{1}{c|}{451045.6963} & \multicolumn{1}{c|}{1.6718} & \multicolumn{1}{c|}{40} & \multicolumn{1}{c|}{49.9517} & \multicolumn{1}{c|}{31.7494} & 120 & \multicolumn{1}{c|}{451093.0293} & \multicolumn{1}{c|}{119.9938} & 92 & \multicolumn{1}{c|}{2.6186} & \multicolumn{1}{c|}{-86.5726} & 68 \\ \hline
\textbf{IVR-IVRU} & \multicolumn{1}{c|}{451045.6961} & \multicolumn{1}{c|}{2.7771} & \multicolumn{1}{c|}{45} & \multicolumn{1}{c|}{49.9517} & \multicolumn{1}{c|}{39.9716} & 170 & \multicolumn{1}{c|}{451093.0291} & \multicolumn{1}{c|}{200.0407} & 140 & \multicolumn{1}{c|}{2.6186} & \multicolumn{1}{c|}{-157.2920} & 75 \\ \hline
\textbf{NFA-NFAU} & \multicolumn{1}{c|}{436385.0531} & \multicolumn{1}{c|}{0.1098} & \multicolumn{1}{c|}{30} & \multicolumn{1}{c|}{49.9517} & \multicolumn{1}{c|}{0.4596} & 35 & \multicolumn{1}{c|}{436434.5157} & \multicolumn{1}{c|}{2.4760} & 31 & \multicolumn{1}{c|}{0.4891} & \multicolumn{1}{c|}{-1.9066} & 34 \\ \hline
\end{tabular}
\end{table*}

\subsubsection{Cost, Runtime, and Iteration Analyses}

The key metrics used to evaluate the performance of PMITD against the \textit{independent} scenarios are the objective cost (in \$/hr), the runtime (in seconds), and the number of IPOPT iterations. It should be noted that for test cases where there are multiple distribution systems, the \textbf{Iterations} for the \textbf{PMD} is the summation of IPOPT iterations for all the independent distribution system problems. 

Table \ref{tab:results_case1} shows the results of the analyzed scenario for test case 1 comparing the integrated formulation of the problem solved using PMITD against the \textit{independent} formulations. As observed, solving the integrated problem formulation provides a better cost than the independent formulation due to the seamless coordination of the transmission and distribution systems generation. In terms of IPOPT iterations, all formulations (e.g., AC-polar, AC-rect, etc.) require a lower total number of iterations to solve the ITD problem. In terms of runtime, only the AC-rect and transportation formulations take less time to solve when the integrated problem is used. 

Tables \ref{tab:results_case2}, \ref{tab:results_case3}, and \ref{tab:results_case4} showcase the results for the other three test cases presented. These test cases are larger and more representative of real systems. As observed in these results, cost is consistently being saved as a result of solving the ITD OPF problem. Also, note how the highest cost savings appear in test case 3, where the DG in the distribution system is rated at 40 MW; thus, as more total distributed generation is available in active distribution systems, solving the ITD problem will become more valuable. Another important detail to note is how the integrated formulation problem takes more runtime to be solved since the overall problem is bigger; however, in terms of total IPOPT iterations, the ITD formulation seems to need a lower number of total iterations when compared to the \textit{independent} test cases (where iterations are added from all the individual problems needed to be solved).

From these results, it is clear that the use of an ITD formulation provides advantages when active distribution systems are considered. Nonetheless, the use of the ITD formulation does not only benefits cost savings. The ITD formulation provides other operational advantages related to the way T\&D systems interact with each other. This is due to the fact that power and voltage values at the distribution system boundary buses match exactly the power and voltage values at the transmission system boundary buses, thus, allowing operators to have a global view of the overall system without the need for variables approximations.

\subsection{Time-Series \& DER integration}

In this experiment, a 24 hours OPFITD problem is solved for a T\&D system that contains multiple DERs (e.g., DGs, PV systems, and energy storage). The primary objective of this experiment is to showcase the PMITD capabilities related to time-series (multinetwork) optimization and DER integration.

\subsubsection{Test Cases}

\begin{figure}[h]
\centering
\includegraphics[width = 0.4\textwidth]{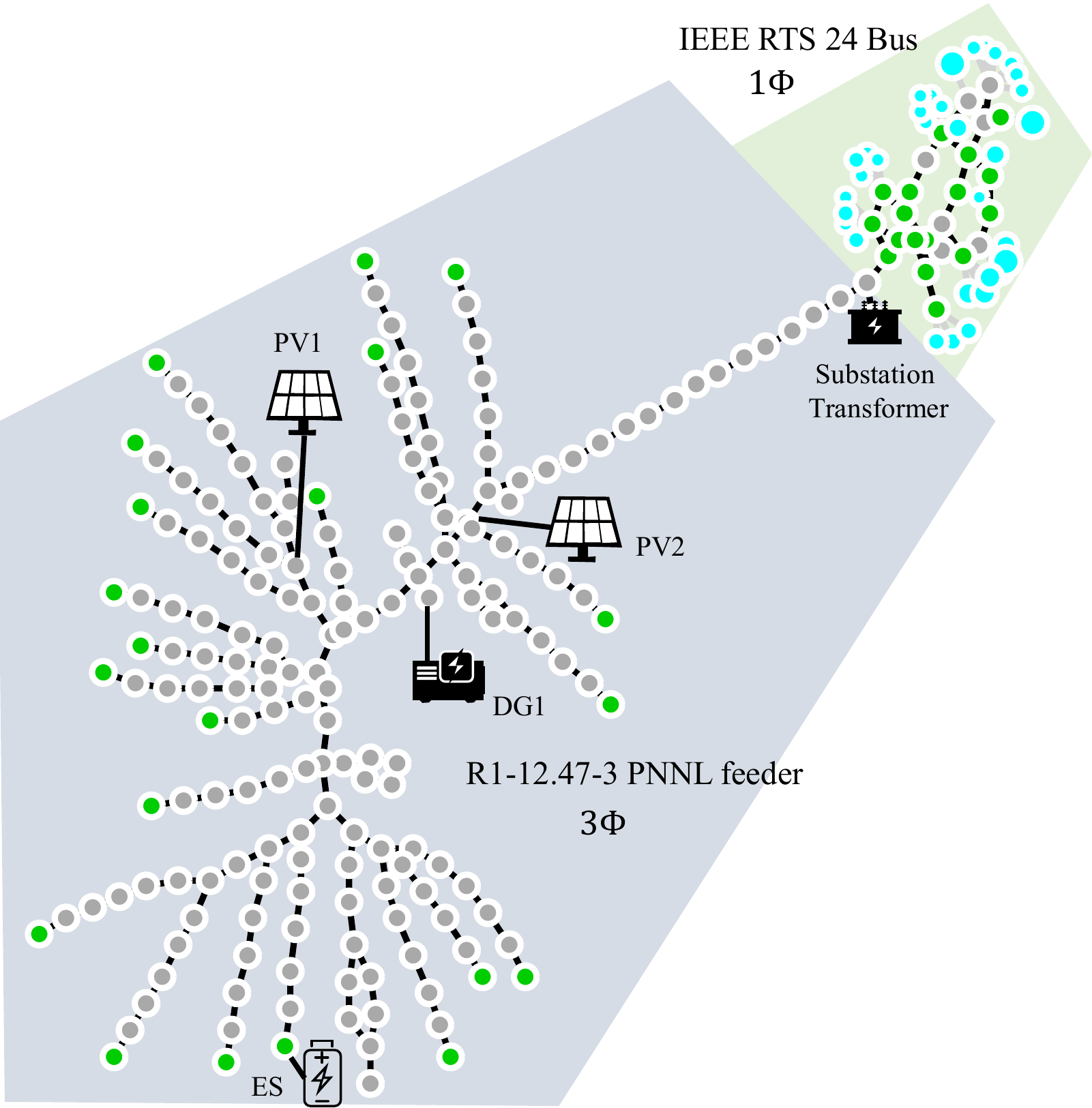}
\caption{\label{fig:dergraph} IEEE RTS 24 - R1-12.47-3 PNNL feeder graph. `Green' nodes represent nodes with loads, `grey' nodes represent connecting nodes, and `light blue' nodes represent transmission generators.}
\end{figure}

The test case used for this experiment is based on modified versions of the IEEE RTS 24 bus system (transmission) and the Feeder 3: R1-12.47-3 (distribution) from PNNL \cite{schneider2008modern}. The load in bus \#4 of the transmission is replaced with a R1-12.47-3 distribution system that contains one distributed generator rated at 300 kW in node \#16, one energy storage (ES) system rated at 200 kW/200 kWh connected with load \#18, and two PV systems rated at 200 and 300 kW connected in nodes \#7 and \#21, respectively. The load profiles for the all the loads connected at the transmission system level are pseudo-randomly generated based on residential and community load profiles that come from \cite{baseloadprofs}. Fig. \ref{fig:dergraph} shows the graph for the described ITD network. Note that for this time-series case scenario, stored energy (SE) constraints for the ES system are added to the problem specification \cite{coffrin2018powermodels}. These constraints handle the temporally-coupled dynamics of the ES system's charging and discharging operations, thus linking the multinetwork optimization between time steps.

The number of nodes and edges for the test case are 192 and 199, respectively. Since this is a 24 hours time-series scenario, the total number of nodes and edges for the problem solved are 4,608 and 4,776. The total number of variables (as given by IPOPT) is 60,750. The problem is solved using the ACR-ACRU ITD formulation.

\subsubsection{Time-Series Results}

\begin{figure}[h]
\centering
\includegraphics[width = 0.45\textwidth]{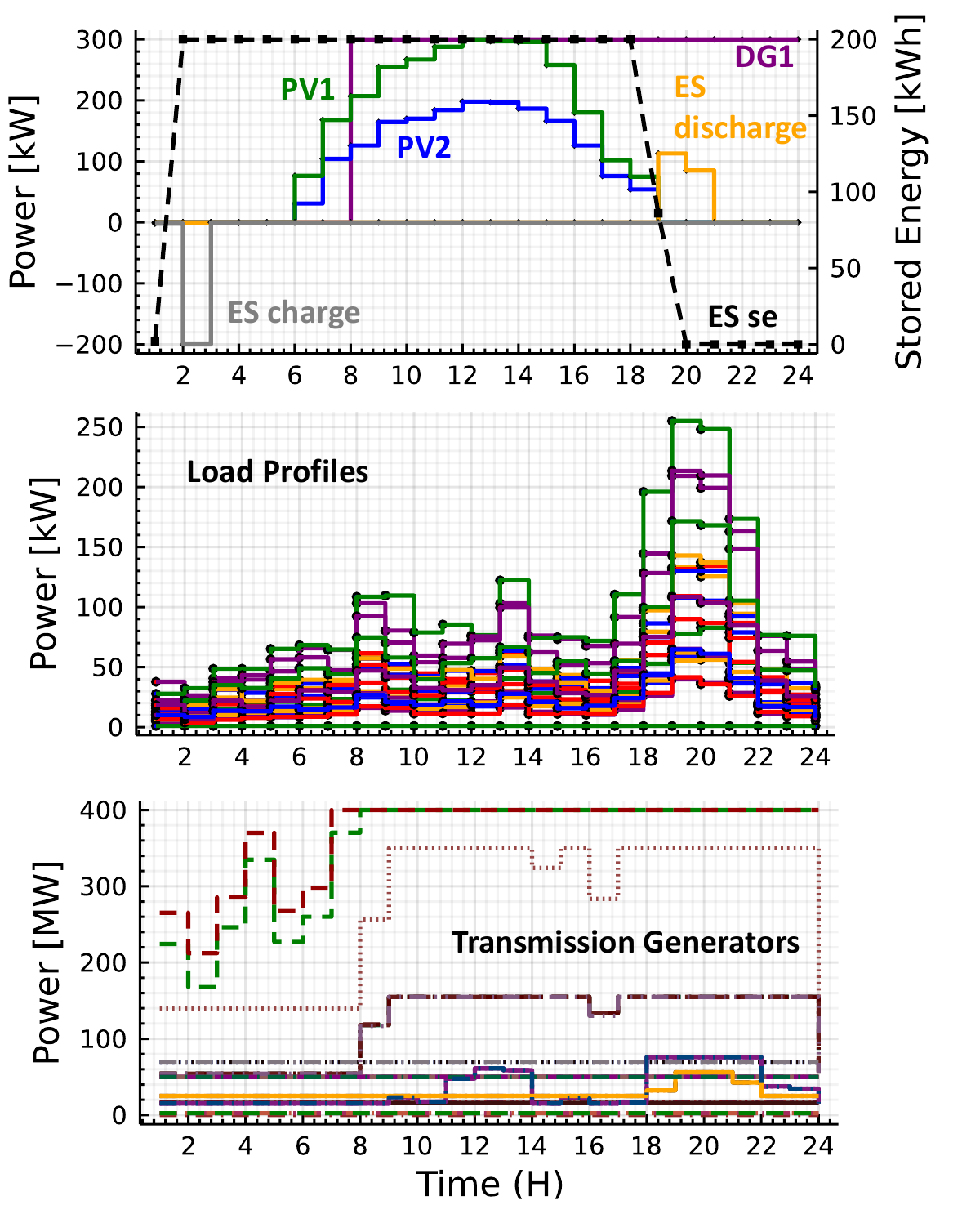}
\caption{\label{fig:derplots1}  24 hours time-series (multinetwork) generation and load profiles. The first graph shows the power dispatch of the DERs. The second graph shows the load profiles of all loads in the distribution system. The third graph shows the power dispatch from the transmission system generators.}
\end{figure}

\begin{figure}[h]
\centering
\includegraphics[width = 0.45\textwidth]{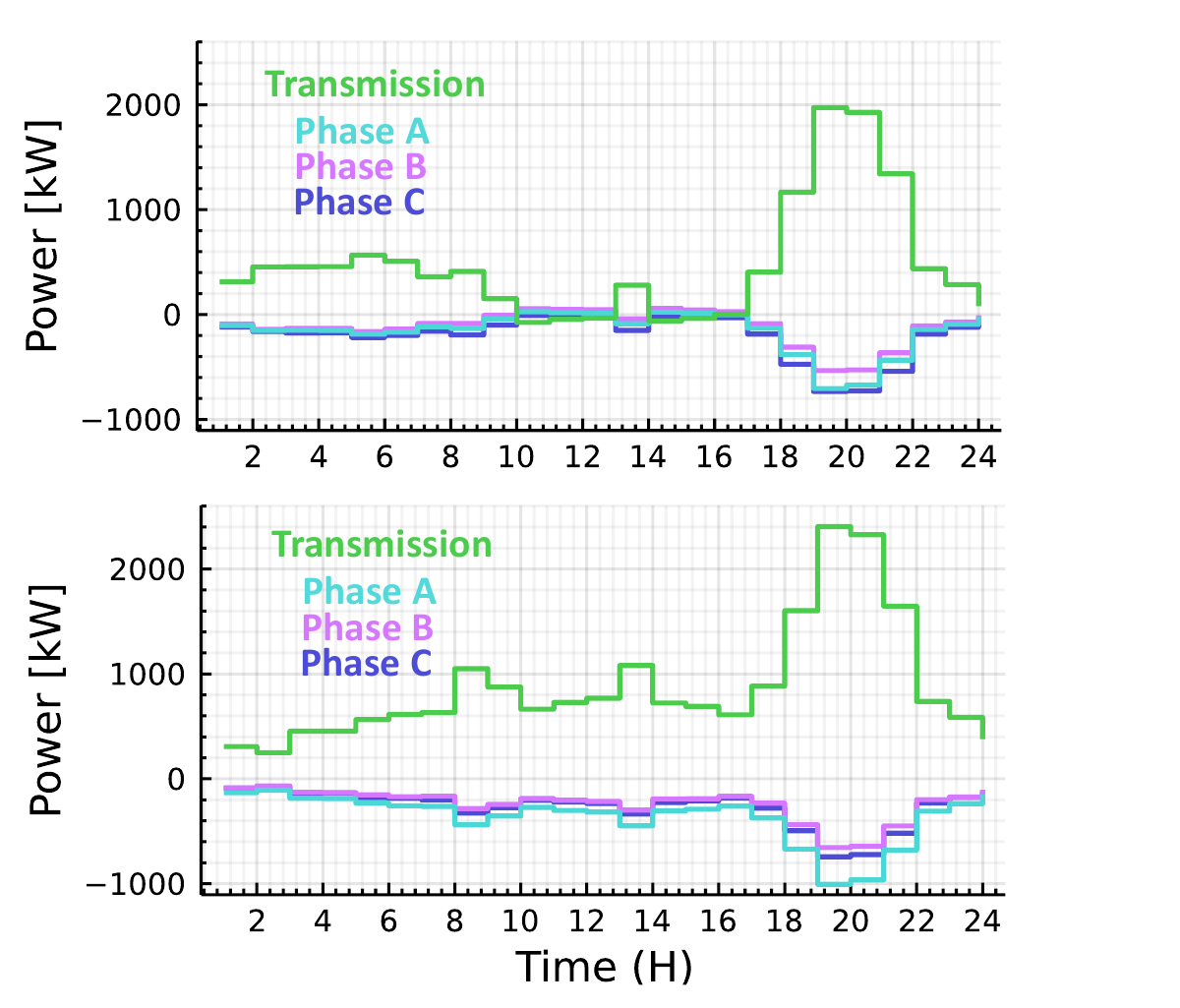}
\caption{\label{fig:derplots2} 24 hours time-series boundary active power flows. The top graph shows the boundary power flows for the DER case. The bottom graph shows the boundary power flows for the same case without DERs. The power flows in phase A, B, C belong to the distribution system. The negative sign of the power flowing to the distribution system is by convention.}
\end{figure}

Fig. \ref{fig:derplots1} shows the 24 hours time-series generation and load profiles for the evaluated test case. The first graph shows the generation coming from the DERs in the feeder and the stored energy (SE) of the energy storage. The second graph shows the load profiles for the loads in the feeder (i.e., distribution system). The third graph shows the generation coming from the transmission generators. Fig. \ref{fig:derplots2} shows the 24 hours time-series boundary active power flows for the evaluated test case. The top graph shows the boundary power flows for the case with DERs. The bottom graph shows the boundary power flows for the same case but without DERs. The cost of the scenario with DERs is \$1,184,660, with 72 iterations and 40.8 seconds. The cost for the scenario without DERs is \$1,184,800, with 36 iterations and 19.9 seconds.

\subsection{Scalability}
\label{sect:experiment3}

\begin{figure}[h]
\centering
\includegraphics[width = 0.48\textwidth]{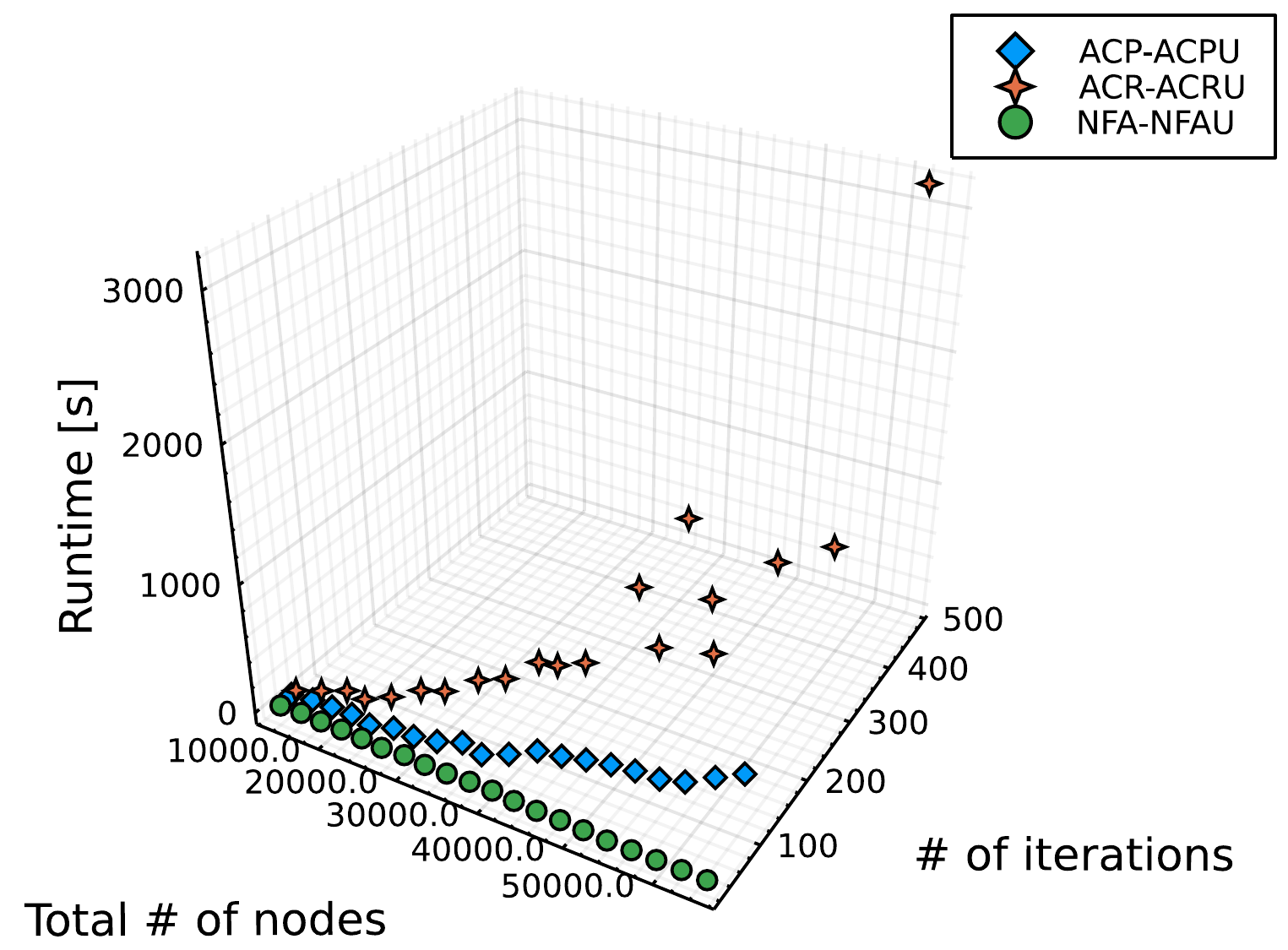}
\caption{\label{fig:scalability} Solver runtime and number of iterations for specific total number of nodes based on particular problem formulation (ACP-ACPU, ACR-ACRU, or NFA-NFAU).}
\end{figure}

In order to assess the scalability of PMITD, a controlled experiment, where the total number of nodes of the ITD problem is gradually increased, is performed. For this experiment, Case 4 from \ref{sect:experiment1} is used as base test case. The only difference with Case 4 is that, for this experiment, the number of distribution systems (IEEE LVTestCase) connected to the IEEE PGLib 500-bus transmission system is gradually increased from 1 to 20, and these distribution systems are connected at load buses \#1 to \#21 (skipping bus \#20). These tests, similar to all other tests run in this section, are performed in machine with a CPU clocked at 2.80 Ghz and 128 GB of RAM. Fig. \ref{fig:scalability} presents a 3-D graph that shows the solver runtime and solver iterations for a specific total \# of nodes in a particular run (1-20) based on a particular problem formulation (ACP-ACPU, ACR-ACRU, or NFA-NFAU). These plots provide a sense of the scalability and limitations of the PMITD  package. These limits are largely related to the available computational power and/or the complexity of the nonlinear nonconvex formulation model used. Note that there is an outlier in the ACR-ACRU case. This outlier is related to the scenario where 19 distribution system are connected to the transmission system; IPOPT has problems converging to the optimal solution.

\section{Conclusion}
\label{sect:conclusion}

In this article, the design and implementation of an open-source framework for solving ITD power network optimization problems is presented. The toolbox is designed to be a complement to \textit{\pmjl} and \textit{\pmdjl}, supporting a diverse set of formulations enabling the co-optimization of transmission and distribution networks. The framework is based on a centralized ITD problem formulation where boundary variables and constraints are defined to allow single-phase transmission systems to be interfaced with unbalanced three-phase distribution systems. The value and performance of the proposed toolbox have been evaluated using a 24 hours time-series test case and using test cases that evaluate the ITD optimization proposed against the independent version of the test cases, where the transmission and distribution systems are solved independently, using a variety of IEEE transmission models and distribution feeders.

The proposed package is designed to be a foundational tool that will enable the speedy development of novel co-optimization algorithms, problem specifications, and formulations that could improve the state-of-the-art. The development of \textit{\pmitdjl} is an ongoing effort with new formulations based on novel approximations and relaxations, such as second-order cone (SOC) and semidefinite programming (SDP), being added, and other decomposition methods that would allow the parallel computation of large-scale problems being developed.

\section*{Acknowledgments}
\label{sect:acknowledgements}
This work was performed with the support of the U.S. Department of Energy (DOE) Office of Electricity (OE) Advanced Grid Modeling (AGM) Research Program under program manager Ali Ghassemian. We gratefully acknowledge Ali's support of this work. The research work conducted at Los Alamos National Laboratory is done under the auspices of the National Nuclear Security Administration of the U.S. Department of Energy under Contract No. 89233218CNA000001.

\IEEEpeerreviewmaketitle

\bibliographystyle{IEEEtran}
\bibliography{biblio}

\placetextbox{0.6}{0.15}{LA-UR-22-22085}

\newpage

\vfill

\end{document}